\documentclass[11pt]{article}

\usepackage{setspace}
\usepackage{cancel}
\usepackage{simplewick}
\usepackage{multirow}
\usepackage{comment}
\usepackage{bbold}
\usepackage{graphicx}
\usepackage{caption}
\usepackage{amsmath}
\usepackage{array}
\usepackage{subcaption}
\usepackage{epstopdf}
\usepackage{enumerate}
\usepackage{cite}
\usepackage{youngtab}
\usepackage{tensor}
\usepackage{slashed}
\usepackage[aligntableaux=center]{ytableau}
\usepackage[utf8]{inputenc}
\usepackage{rotating}
\usepackage{bigfoot}
\usepackage{xfrac}

\usepackage{booktabs}
\usepackage{siunitx}
	
\usepackage{color, colortbl}
	
\definecolor{Gray}{gray}{0.9}

\usepackage[
colorlinks=true,
linkcolor=black,
urlcolor=blue,
filecolor=black,
citecolor=red,
]{hyperref}

\numberwithin{equation}{section}

\def \( {\left(}
\def \) {\right)}
\def \< {\left<}
\def \> {\right>}

\newcommand{\be}{\begin{equation}} \newcommand{\ee}{\end{equation}}
\newcommand{\bea}{\begin{eqnarray}}  \newcommand{\eea}{\end{eqnarray}}
\newcommand{\nn}{\nonumber}

\usepackage{fullpage}
\usepackage{verbatim}
\usepackage{bm,latexsym,amsmath,amssymb,amsfonts,mathrsfs,amsthm,mathtools}
\usepackage{bbm} 
\usepackage{graphicx}
\usepackage{color}
\input{colordvi.tex}
\usepackage{latexsym}
\usepackage{epsf}
\usepackage{slashed} 
\usepackage{array} 
\usepackage{booktabs}
\usepackage{environ} 

\renewcommand{\d}{\textrm{d}}

\DeclarePairedDelimiter\bra{\langle}{\rvert}
\DeclarePairedDelimiter\ket{\lvert}{\rangle}
\DeclarePairedDelimiterX\braket[2]{\langle}{\rangle}{#1 \delimsize\vert #2}

\def\calb         {{\cal B}}

\def\calf         {{\cal F}}

\def\calh         {{\cal H}}

\def\call         {{\cal L}}
\def\calm         {{\cal M}}
\def\caln         {{\cal N}}
\def\calo         {{\cal O}}

\def\caly         {{\cal Y}}

\newcommand{\del}{\partial}

  \newcommand{\ads}{\text{AdS}}

\renewcommand{\d}{\mathrm{d}}

\NewEnviron{NORMAL}{%
    \scalebox{2}{$\BODY$} 
} 
 
\NewEnviron{HUGE}{%
    \scalebox{5}{$\BODY$} 
} 

\begin{document}
	\interfootnotelinepenalty=10000 

	\vspace*{-1.5cm}
	\begin{flushright}    
		{\small
		}
	\end{flushright}
	
	\vspace{1.8cm}
	\begin{center}        
	\LARGE

		Self-Binding Energies in AdS
		
	\end{center}
	
	\vspace{0.7cm}
	\begin{center}        
		{\large Stefano Andriolo, Marco Michel, Eran Palti }
	\end{center}
	
	\vspace{0.15cm}
	\begin{center}        
		\emph{Department of Physics, Ben-Gurion University of the Negev, Be'er-Sheva 84105, Israel}\\[.3cm]
		\emph{}\\[.4cm]
		
		e-mails: \tt stefanoa@post.bgu.ac.il, \tt michelma@post.bgu.ac.il, \tt palti@bgu.ac.il
	\end{center}
	
	\vspace{1.5cm}
	
	
	\begin{abstract}
	\noindent The Positive Binding Conjecture is a proposed formulation of the Weak Gravity Conjecture appropriate to Anti de-Sitter (AdS) space. It proposes that in a consistent gravitational theory, with a $U(1)$ gauge symmetry, there must exist a charged particle with non-negative self-binding energy. In order to formulate this as a constraint on a given effective theory, we calculate the self-binding energy for a charged particle in AdS$_4$ and AdS$_5$. In particular, we allow it to couple to an additional scalar field of arbitrary mass. Unlike the flat-space case, even when the scalar field is massive it contributes significantly to the binding energy, and therefore is an essential component of the conjecture. In AdS$_5$, we give analytic expressions for the self-binding energy for the cases when the scalar field is massless and when it saturates the Breitenlohner-Freedman (BF) bound, and in AdS$_4$ when it is massless. We show that the massless case reproduces the flat-space expressions in the large AdS radius limit, and that both analytic cases lead to vanishing total self-binding energy for BPS particles in example supersymmetric models. For other masses of the scalar we give numerical expressions for its contribution to the self-binding energy.
	\end{abstract}
	
	\thispagestyle{empty}
	\clearpage
	
	\tableofcontents
	
	\setcounter{page}{1}
	
	
	
	\section{Introduction}
	\label{sec:intro}
		
	The Weak Gravity Conjecture \cite{Arkani-Hamed:2006emk} plays a central role in the Swampland programme \cite{Vafa:2005ui} (see \cite{Palti:2019pca, vanBeest:2021lhn,Harlow:2022gzl} for reviews). A particular formulation of the conjecture in flat space is the Repulsive Force Conjecture \cite{Palti:2017elp,Heidenreich:2019zkl}, which proposes that a consistent gravitational theory, with a $U(1)$ gauge symmetry, must have a self-repulsive charged particle: so a particle which would repel an identical copy of itself. This formulation constrains all interactions in an effective theory which contribute to the self-force of the particle. In particular, any massless scalar fields which couple to the particle contribute to the self-force (even at long range) and therefore appear in the constraints \cite{Palti:2017elp}. 

If we consider Anti-de Sitter Space (AdS), then a proposed formulation is the Positive Binding Conjecture in \cite{Aharony:2021mpc}.\footnote{For alternative approaches to AdS see \cite{Nakayama:2015hga,Montero:2016tif,Montero:2018fns} for example.} The conjecture proposes that the theory must have a charged particle which has a positive (or vanishing) self-binding energy. This formulation is also particularly interesting because it maps to certain convexity properties of charged operators in the holographically dual CFTs \cite{Aharony:2021mpc}.\footnote{See, for example, \cite{Aalsma:2021qga,Watanabe:2022htq,Antipin:2021rsh,Moser:2021bes,Dupuis:2021flq,Palti:2022unw} for work on this topic.} Demanding positive self-binding energy places constraints on the effective theory. In this paper we determine what those constraints are by calculating the self-binding energy for a charged particle in terms of its couplings. We consider effective theories in AdS$_5$ and AdS$_4$. An expression for the self-binding energy in the case when the particle couples to gravity, the $U(1)$ photon, and has some quartic contact terms was calculated in \cite{Fitzpatrick:2011hh}. We generalise these results to the case when the charged particle couples also to a neutral scalar field (as well as cases with additional quartic terms). The neutral scalar is allowed to have arbitrary mass. However, for a general mass, the expression for the binding energy is very complicated and we evaluate it only numerically. We do find analytic results for special values of the mass. In AdS$_5$ we present analytic results for the cases when it is massless and when it saturates the Breitenlohner-Freedman (BF) bound. In AdS$_4$ we give an analytic expression for the massless case.

Our results contribute towards sharpening proposed quantum gravity constraints on effective theories in AdS. Such theories are interesting in themselves, but also are particularly utilised to study holography. They also move us closer towards being able to quantitatively test the Weak Gravity Conjecture, or more precisely the Positive Binding Conjecture, in string theory constructions of AdS space. 

\subsubsection*{Summary of Results}

The calculation of the self-binding energy is somewhat involved. We therefore present a summary of the results here for convenience. 

We consider a theory in AdS with the following matter spectrum. There is a charged scalar $\phi$, for which we calculate the self-binding energy, which is charged under $N$ $U(1)$ gauge fields. There is also a neutral scalar $\chi$, which couples to $\phi$ and therefore contributes to the binding energy. 
The general action in AdS in $d$-dimensions is
\begin{align}
\nn
S[\phi,A^i,h,\chi] 
&= 
\int \d^dx \sqrt{-g} 
\bigg[ 
\frac{1}{\kappa^2} \left( \frac{R}{2} - \Lambda \right) 
- \sum_{i=1}^N \frac{F_i^2}{4} 
-  |D\phi|^2 
- m^2|\phi|^2 
- V(\phi)   \\
&\qquad\qquad\qquad\qquad
- \frac{1}{2} (\del\chi)^2 
- \frac{M^2}{2}\chi^2
- Y \chi |\phi|^2 
- \delta \chi |\del\phi|^2 
\bigg]
\,.
\label{setup_d}
\end{align}
Here, with $\kappa^{-2}\equiv M_{d}^{d-2}$, with $M_d$ being the associated Planck's constant. The cosmological constant is $\Lambda=-\tfrac{(d-1)(d-2)}{2L^2}$, where $L$ is the AdS radius. There is (an effective) potential containing the following contact terms
\begin{align}
V(\phi) &= 
a |\phi|^4 
+ b |\phi|^2 |\del\phi|^2 
+  c \left( \phi^2 (\del \phi^\dag)^2  +  (\phi^\dag)^2 (\del \phi)^2 \right) 
\,,
\label{V_quartic}
\end{align}
which are the quartic terms contributing to the binding energy. The covariant derivative of the charged scalar includes the gauge couplings $g_i$, and integer charges $q_i$, as
\begin{align}
D_\mu\phi = \del_\mu\phi - i \sum_{i=1}^N g_i q_i A_{i\,\mu} \phi 
\quad 
i=1,\dots,N
\,.
\label{covD}
\end{align}

For this effective action we find that (at weak couplings) the self-binding energy for $\phi$, denoted as $\gamma$,  is composed of a number of contributions
\be
\gamma = \gamma^V + \gamma^{\mathrm{phot}} + \gamma^{\mathrm{grav}} + \gamma^{\mathrm{scal}} \;,
\label{totabind}
\ee
where the labels are naturally associated to the type of contribution. 
It is useful, when presenting the binding energies, to exchange the mass of the charged scalar $m$ for its holographic dual operator dimension $\Delta$, so writing
\be
m^2L^2 = \Delta(\Delta-d+1) \;.
\ee
The self-binding energy contributions from the contact terms, the photon, the graviton and special cases of the scalar, for AdS$_4$ and AdS$_5$ are shown in table 1.\footnote{We also have an analytic expression in AdS$_4$ for the case $M^2L^2=-2$. However, it is very long and complicated and so we refrain from displaying it in the paper.} 
\bgroup
\def\arraystretch{2.4}
\begin{table}[t!]
\begin{center}
\begin{tabular}{| c | c | c |}
\hline
 & AdS$_5$ & AdS$_4$  \\
\hline
$N_\Delta$ & $\sqrt{\frac{\Delta-1}{2\pi^2}}$ & $\sqrt{\frac{\Gamma(\Delta+1)}{2\Delta\Gamma(\Delta-1/2)\pi^{\frac32}}}$ \\
\hline
$\gamma^V$ &
$\frac{\pi^2 N_{\Delta }^4 \left(a
   L^2-b (\Delta -2) \Delta + 2c \Delta^2 
   \right)}{(\Delta -1) (2 \Delta -1) L^4}$
&
$
\frac{\pi^{3/2} N_{\Delta}^4 \Gamma\left(2\Delta-\tfrac{3}{2}\right)\left(2a L^2-b (2\Delta - 3) \Delta + 4\Delta^2 c \right)}{\Gamma(2\Delta)L^3}$ \\
\hline
$\gamma^{\mathrm{phot}}$ &
$\frac{\pi^2 N_{\Delta }^4}{L^2(2 \Delta-1)}  \sum_i g_i^2 q_i^2$ 
&
$2 \pi^{3/2} \frac{N_{\Delta }^4}{L} 
\frac{\Gamma(2\Delta-1/2)}{\Gamma(2\Delta)}  \sum_i g_i^2 q_i^2 $
\\
\hline
$\gamma^{\mathrm{grav}}$ & 
$-\frac{2 \pi ^2 (\Delta -2) \Delta
   ^2 \kappa^2 N_{\Delta }^4}{3
   (\Delta -1) (2 \Delta -1) L^4}$
& 
$- \pi^{3/2} \kappa^2 N^4_\Delta \Delta^2 (2\Delta-3) 
\frac{\Gamma(2\Delta-3/2)}{\Gamma(2\Delta) L^3} $
\\
\hline
 $\gamma^{\mathrm{scal}}_{M^2L^2=-4} $
& $- \frac{Y^2 \pi^2 N_{\Delta }^4 }{8 (\Delta-1)^3}$ & --- \\
\hline
  $\gamma^{\mathrm{scal}}_{M=0} $ & (\ref{be_scalar_massless}) & (\ref{be_scalar_massless_4})\\
\hline
\end{tabular}
\label{tab:bindsum}	
\caption{\emph{Table showing the contributions to the total binding energy, as given by (\ref{totabind}), for the contact terms, the photon and the graviton. The scalar contribution in AdS$_5$, for a special value of the scalar mass saturating the BF bound is also presented. The massless cases are referred to through equations in the main text.  The first row gives the normalization factor $N_\Delta$, which is used in the binding energy expressions. $\Gamma$ denotes the Gamma function.}}
\end{center}
\end{table}
\egroup

It is worth pointing out that we find that the term with the $\delta$ coefficient in (\ref{setup_d}) contributes to the binding energy and also survives the flat space limit, see (\ref{SWGC_5}) and (\ref{flat_limit_4}). It is therefore a new contribution even to the binding energy, or repulsive forces, in flat space (though is typically suppressed as a non-renormalizable operator). 

\medskip
The layout of the paper is as follows. In Section \ref{sec:BE} we introduce the general formalism, following \cite{Fitzpatrick:2011hh}, for calculating the binding energies. In Section \ref{sec:5d} we then calculate the binding energies in AdS$_5$, and in Section \ref{sec:4d} we present the results for the calculation in AdS$_4$. Finally, in Section \ref{sec:tests}, we show that the massless cases reproduce the flat-space expressions of \cite{Palti:2017elp,Lee:2018spm} in the large AdS radius limit, and present two examples of supersymmetric models yielding a vanishing total self-binding energy for BPS particles.

\medskip
Before proceeding, let us clarify some conventions. We parametrize $\ads_d$ of radius $L$ using the following global conformal coordinates:\footnote{This is just a more convenient parametrization of 
\begin{align}
\label{AdS_global_alt}
\d s^2 =  \frac{1}{\cos^2(r/L)} \left(- \d t^2 + \d r^2 + L^2 \sin^2(r/L) \d\Omega_{d-2}^2 \right) 
\,,
\end{align}
obtained using $y=\cos(r/L)$ and redefining $t$. In these coordinates it is easier to understand the flat limit $L\to\infty$.}
\begin{align}
\label{AdS_global}
\d s^2 = \frac{L^2}{y^2} \left(-\d t^2 + \frac{\d y^2}{1-y^2} + (1-y^2) \d\Omega_{d-2}^2 \right) 
\,,
\end{align}
where $y=0$ corresponds to the boundary $\mathbb{R}\times S^{d-2}$ and $y=1$ is the $\ads$ center. Coordinates $(\theta_1,\dots,\theta_{d-3},\varphi)$ parametrize $S^{d-2}$. The flat spacetime limit is obtained by taking $L\to\infty$. Throughout the paper, we work setting $L=1$ and restore the appropriate powers of $L$ as needed.

	\section{Binding energy from effective potentials}
	\label{sec:BE}

Consider a theory for fields $\Phi=(\phi,A_{i \mu},h_{\mu\nu},\dots)$ around an $\ads_d$ vacuum with effective action $S[\Phi]$. In particular, let $\phi$ be a scalar charged under some photons $A_{i \mu}$. Suppose we know the spectrum of this theory. Let $\ket{\phi},\ket{\phi\phi}$ the single- and two-particle $\phi$ states of lowest energy, and let their energy be $E_{\phi},E_{\phi\phi}$, respectively. We want to compute the \emph{self-binding energy} $\gamma$ of $\ket{\phi\phi}$, defined as the difference between $E_{\phi\phi}$ and $E_{\phi}$. In the Hamiltonian formalism, this is
\begin{align}
\label{gamma_def}
\gamma \equiv E_{\phi\phi} - 2 E_{\phi} = \bra{\phi\phi}H\ket{\phi\phi} - 2\bra{\phi}H\ket{\phi}
\,,
\end{align}
where $H$ is the Hamiltonian of the system $S[\Phi]$. 
It is thus clear that $\gamma$ depends on interactions, in absence of which $\gamma=0$ trivially. 

\begin{figure}[t!]
	\centering
	\includegraphics[scale=0.52]{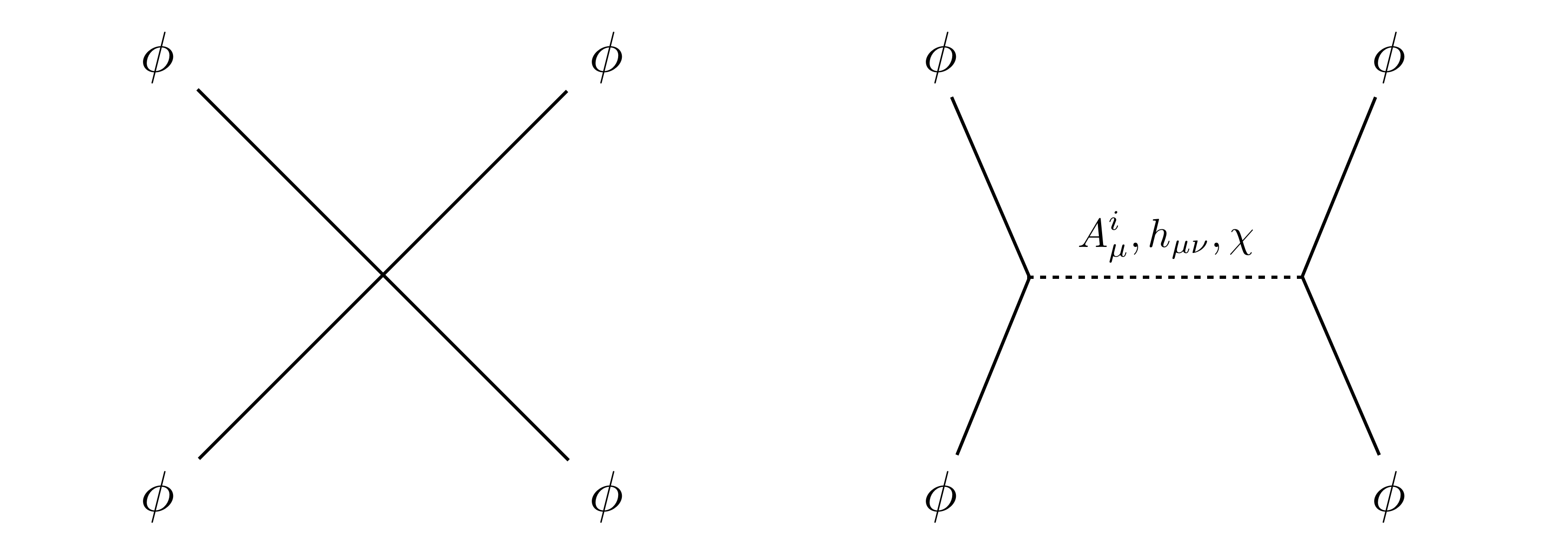}
	\caption{\emph{Feynman diagrams for four-point interactions at tree-level: point-like, as $\phi^4,\phi^2(\del\phi)^2,...$ (left), or mediated by gauge bosons $A^i_\mu$ (e.g., photons), graviton or neutral scalar $\chi$ (right).}}
		\label{fig:graphs}
\end{figure}

\medskip
We are interested in computing the binding energy \eqref{gamma_def} perturbatively in the couplings, at tree level. The contributions to the binding energy come from quartic contact terms and from an exchange of another field, as illustrated in Fig.~\ref{fig:graphs}. 
By treating interactions as small perturbations of the free theory for $\phi$, one can compute $\gamma$ with the aid of perturbation theory in the Hamiltonian formalism \cite{Fitzpatrick:2010zm,Fitzpatrick:2011hh}. 
The Hamiltonian is $H=H_{\rm free}+\delta H$, where $\delta H$ encodes small interactions, and the operator $\phi$, as well as states $\ket{\phi},\ket{\phi\phi}$ in \eqref{gamma_def}, are taken in the free theory (see Appendix \ref{app:free} for details on quantization of a free scalar $\phi$).
To compute contributions to $\gamma$ given by field exchanges one needs second order perturbation theory. This is a difficult task, since it requires summing over all possible intermediate states \cite{Fitzpatrick:2010zm}. 
Alternatively, as suggested in \cite{Fitzpatrick:2011hh}, one can classically integrate out all the fields apart from $\phi$, so in our setting $A^i_{\mu},h_{\mu\nu}$ and $\chi$, in order to obtain an effective action for $\phi$. Using this effective theory, $\gamma$ can be computed entirely within first order perturbation theory. 

Let us illustrate how such procedure works by specifying the action $S[\Phi]$.
The simplest tree-level (two-derivative) action $S[\Phi]$ producing diagrams such as those of Fig.~\ref{fig:graphs} is:  
\begin{align}
\nn
S[\phi,A^i,h,\chi] 
&= 
\int \d^dx \sqrt{-g} 
\bigg[ 
\frac{1}{\kappa^2} \left( \frac{R}{2} - \Lambda \right) 
- \sum_{i=1}^N \frac{F_i^2}{4} 
-  |D\phi|^2 
- m^2|\phi|^2 
- V(\phi)   \\
&\qquad\qquad\qquad\qquad
- \frac{1}{2} (\del\chi)^2 
- \frac{M^2}{2}\chi^2
- Y \chi |\phi|^2 
- \delta \chi |\del\phi|^2 
\bigg]
\,,
\label{setup_d}
\end{align}
with $\kappa^{-2}\equiv M_{d}^{d-2}$, cosmological constant $\Lambda=-\tfrac{(d-1)(d-2)}{2L^2}$ and a scalar potential containing the following contact terms
\begin{align}
V(\phi) &= 
a |\phi|^4 
+ b |\phi|^2 |\del\phi|^2 
+  c \left( \phi^2 (\del \phi^\dag)^2  +  (\phi^\dag)^2 (\del \phi)^2 \right) 
\,.
\label{V_quartic}
\end{align}
Notice that in the interactions of \eqref{V_quartic} we are taking $D\to\del$ since neglected terms in the covariant derivative yield interaction vertices that are not relevant to our analysis.\footnote{\label{foot:extraterms} One may also wonder why we have not included in \eqref{setup_d} scalar couplings of the form $\del^\mu\chi (\phi^\dag\del_\mu\phi \pm \text{h.c.} )$, with real and imaginary coefficients respectively. Remembering that $\phi$ is free, one can integrate by parts and use the equation of motion for $\phi$ to find that (up to total derivatives) the term with minus sign vanishes, and the term with plus sign yields couplings $\chi |\phi|^2$ and $\chi |\del\phi|^2$, both of which we have already included.}

The complex scalar $\phi$ is charged under $U(1)^{N}$, with charges $q_i$'s, via the standard covariant derivative
\begin{align}
D_\mu\phi = \del_\mu\phi - i \sum_{i=1}^N g_i q_i A_{i\,\mu} \phi 
\quad 
i=1,\dots,N
\,,
\label{covD}
\end{align}
where $g_i$ is the coupling constant of each $U(1)$. 
In \eqref{setup_d}, the following expansion around the AdS background is assumed, $g_{\mu\nu} \to g_{\mu\nu} + \kappa h_{\mu\nu}$, such that the graviton $h_{\mu\nu}$ has a canonical kinetic term. 
All fields have the same mass dimension, $[\phi]=[\chi]=[A^i]=[h]=\tfrac{d-2}{2}$, while $g_i,Y,\delta,a,b$ have dimensions $2-\tfrac{d}{2},3-\tfrac{d}{2}, 1-\tfrac{d}{2},4-d,2-d$ respectively ($q_i$'s are dimensionless).

\medskip
We classically integrate out fluctuations $A^i_{\mu},h_{\mu\nu},\chi$ by computing their linearized equations of motion and plugging their solutions back into the action \eqref{setup_d}. As a result, one obtains the tree-level effective action 
\begin{gather}
\nn
S_{\rm eff} = S_{\rm free} - \int \d^dx \sqrt{-g} ~ V_{\rm eff} [\phi,\phi^\dag]
\,, \\
\nn
S_{\rm free} = 
\int \d^d x \sqrt{-g} 
\left[ 
- |\del\phi|^2 
- m^2|\phi|^2
\right]
\,,\\
V_{\rm eff} [\phi,\phi^\dag] 
= V [\phi,\phi^\dag] 
+ V_{\rm eff,phot} [\phi,\phi^\dag]  
+ V_{\rm eff,grav} [\phi,\phi^\dag]  
+ V_{\rm eff,scal} [\phi,\phi^\dag]  
\,,
\label{S_eff_implicit}
\end{gather}
where the effective potential contains all contact terms in $V$ \eqref{V_quartic} and the ``new contact terms" $V_{\rm eff,phot}, V_{\rm eff,grav}, V_{\rm eff,scal}$ arising from the integration procedure (i.e., field exchanges). 
All quantities in $V_{\rm eff}$ must be regarded as non-local functionals of $\phi,\phi^\dag$. 

Using the standard definitions of momentum conjugate for $\phi$ one gets 
$\Pi_{\phi}= - g^{tt}\del_t\phi^\dag-\tfrac{\del V_{\rm eff}}{\del(\del_t\phi)}$ and 
$\Pi_{\phi^{\dag}}= - g^{tt}\del_t\phi-\tfrac{\del V_{\rm eff}}{\del(\del_t\phi^\dag)}$, and thus the Hamiltonian density 
$\calh = \Pi_{\phi}\del_t\phi + \Pi_{\phi^\dag}\del_t\phi^\dag + |\del\phi|^2 + m^2 |\phi|^2 + V_{\rm eff}$ is
\begin{align}
\calh = - g_{tt} \Pi_{\phi} \Pi_{\phi^\dag} + |\vec\nabla\phi|^2 + m^2 |\phi|^2 + V_{\rm eff}
= \calh_{\rm free}+ V_{\rm eff}
\,,
\end{align}
where the free Hamiltonian density is given by \eqref{Hdensity_free}. 
Therefore, the Hamiltonian of the system, $H=\int \d^{d-1}x \sqrt{-g} \calh$, is simply
\begin{align}
\label{H}
H = H_{\rm free} + \delta H_{\rm eff}
\,, \qquad
\delta H_{\rm eff} = \int \d^{d-1}x \sqrt{-g} ~ V_{\rm eff} 
\,,
\end{align}
where the integration is over a spatial surface at fixed $t$. 
Remembering that $\bra{\phi\phi} H_{\rm free} \ket{\phi\phi}= 2 E^{\rm free}_{\phi}$, one can use first order perturbation theory to obtain $E_{\phi\phi}$: 
\begin{align}
E_{\phi\phi} 
& \equiv \bra{\phi\phi} H \ket{\phi\phi} 
= 2 E^{\rm free}_{\phi} + \bra{\phi\phi} \delta H_{\rm eff} \ket{\phi\phi} 
= 2 E_{\phi} + \bra{\phi\phi} \delta H^{\rm quartic}_{\rm eff} \ket{\phi\phi} 
\,,
\end{align}
where we used that the quadratic terms in $\delta H_{\rm eff}$ correct the energy of the single-particle state $E_{\phi}=E^{\rm free}_{\phi} + \bra{\phi} \delta H^{\rm quadratic}_{\rm eff} \ket{\phi}$. As expected, the binding energy \eqref{gamma_def} is purely determined by quartic terms in $V_{\rm eff}$ generated by diagrams of Fig.~\ref{fig:graphs}. 
Using \eqref{H} and recalling the following well-known facts regarding quantization of a free $\phi$ in AdS (see Appendix \ref{app:free} for details):
\begin{align}
& \phi = \sum_{nlJ} \big(a_{nlJ} \psi_{nlJ}(x) + b^\dag_{nlJ} \psi^*_{nlJ}(x) \big)
\,, 
\quad 
[b_0 , b_0^\dagger] = 1
\,,
\quad
\ket*{\phi\phi} = \frac{1}{\sqrt{2}} b_0^\dag b_0^\dag \ket*{0} 
\,,
\end{align} 
one finds 
\begin{align}
\nn
\gamma 
&= \int \d^{d-1}x \sqrt{-g} \bra*{\phi\phi} V_{\rm eff, quartic} \ket*{\phi\phi}  
\\
\nn
&= 2 \int \d^{d-1}x \sqrt{-g} \left. V_{\rm eff, quartic}[\phi,\phi^\dag]\right|_{b_0^\dag b_0^\dag b_0 b_0}
\\
\nn
&= 2 \int \d^{d-1}x \sqrt{-g} ~ V_{\rm eff, quartic}[\phi(x)=\psi^*_0(x),\phi^\dag(x)=\psi_0(x)]
\\
& \equiv 2 \int_{S^{d-2}} \d\theta_1\cdots\d\theta_{d-3}\d\varphi \int_0^1 \d y \sqrt{-g} ~ V_{\rm eff, quartic}[\psi^*_0(x),\psi_0(x)]
\,,
\label{be_implicit}
\end{align}
where $V_{\rm eff}$ is normally ordered and $\psi_0(x)$ is the wavefunction with lowest energy. In other words, we need to extract $b_0^\dag b_0^\dag b_0 b_0$ from $V_{\rm eff, quartic}$ in the first line of \eqref{be_implicit}, and we do so by taking $\psi^*_0 b^\dag_0$ in $\phi$ and $\psi_0 b_0$ in $\phi^\dag$. 
By commuting operators, we are left with twice $V_{\rm eff, quartic}[\phi=\psi^*_0,\phi^\dag=\psi_0]$, which is just a functional of $\psi_0(x)$ and can be integrated over space. In the next Sections we will explicitly compute all binding energy contributions given by $V,V_{\rm eff,phot}, V_{\rm eff,grav}, V_{\rm eff,scal}$ in \eqref{be_implicit}. We do so for $\ads_5$ in Section \ref{sec:5d} and for $\ads_4$ in Section \ref{sec:4d}. 

\section{Binding energy in $\ads_5$}
\label{sec:5d}

The $d=5$ case is an extension of the work \cite{Fitzpatrick:2011hh} to several photons, additional contact terms (associated to the parameter $c$), and most importantly, the inclusion of neutral scalar couplings. We start by neglecting $\chi$ and compute $\gamma$ contributions given by $V$, photon and gravition exchanges in Sections \ref{sec:V}, \ref{sec:photons} and \ref{sec:graviton} respectively. Then, in Section \ref{sec:scalars} we extend this procedure to $\chi$.


It is worth mentioning that the perturbation parameters are:
\begin{align}
\label{small_int5d}
\frac{g_i}{L^{1/2}}\ll 1
\,, \quad
\frac{\kappa}{L^{3/2}}\ll 1
\,, \quad
\frac{|a|}{L}\ll 1
\,, \quad
\frac{|b|}{L^3}\ll 1
\,, \quad
\frac{|c|}{L^3}\ll 1
\,, \quad
|Y|L^{1/2}\ll 1
\,, \quad
\frac{|\delta|}{L^{3/2}}\ll 1
\,.
\end{align}
We find that the leading contributions to the binding energy are at $\calo\left(\kappa^2 L^{-3}\right)$, $\calo\left(g_i^2 L^{-1}\right)$, $\calo\left(a L^{-1}\right)$, $\calo\left(b L^{-3} \right)$, $\calo\left(c L^{-3} \right)$, $\calo\left(Y^2 L\right)$ and $\calo\left(\delta^2 L^{-3} \right)$.

\subsection{Contribution from contact terms in $V$}
\label{sec:V}

The contribution to $\gamma$ given by $V$ in \eqref{V_quartic} is given by four types of contact terms. 
Plugging $V$ into \eqref{be_implicit}, using $\phi\to\psi^*_0, \phi^\dag\to\psi_0$ where (see \eqref{app:psi0})
\begin{align}
\psi^*_0 =  N_\Delta e^{it\Delta} y^\Delta
\,, \quad 
N_\Delta = \sqrt{\frac{\Delta-1}{2\pi^2}}
\,, \quad
m^2 = \Delta(\Delta-4)
\,,
\end{align}
and performing the integral, one gets:
\begin{align}
\label{be_quartic}
&\gamma^{V} = 
\frac{\pi^2 N_{\Delta }^4 \left(a
   L^2-b (\Delta -2) \Delta + 2c \Delta^2 
   \right)}{(\Delta -1) (2 \Delta -1) L^4} 
\,,
\end{align}
where we reintroduced $L$. 
As expected, as long as the unitarity bound $\Delta>1$ is respected (as assumed throughout the paper), the boundary contribution to the integral vanishes.

\subsection{Photon exchanges}
\label{sec:photons}

In order to obtain the photon contribution to the effective potential, $V^{\rm phot}_{\rm eff,quartic}$, we need to integrate out photons explicitly. 
The variational principle in AdS, where variations vanish at the boundary, yields the linearized equation:
\begin{align}
\frac{1}{\sqrt{-g}} \partial_\mu (\sqrt{-g} F_i^{\mu\nu}) 
= g_i q_i J^\nu  
\,, 
\label{EOMs_photons}
\end{align}
where the source 
\begin{align}
\label{J}
J_{\mu} [\phi,\phi^\dag] = 
i \phi^\dag \partial_\mu\phi + h.c. 
\,,
\end{align}
is quadratic in $\phi$. As we are about to show, we can neglect terms of second and higher order in $g_i$ since 
it is enough to solve the equations of motion to linear order in perturbations in order to compute $\gamma$ at leading order. This also means that there is no mixing between photon species and so we can focus on the $i$-th species, eventually summing over $i$ the result. Now, let us conveniently take $A_i=A^{(0)} + g_i q_i A^{(1)}$, where $A^{(0)}$ and $A^{(1)}$ are free and perturbed parts, respectively solving
\begin{align}
\label{EOMs_photon_free}
& \frac{1}{\sqrt{-g}} \partial_\mu (\sqrt{-g} F^{(0) \mu\nu}) = 0  
\,, \quad
F^{(0)}=\d A^{(0)}
\,, \\
& \frac{1}{\sqrt{-g}} \partial_\mu (\sqrt{-g} F^{(1) \mu\nu}) =  J^\nu  
\,, \quad
F^{(1)}=\d A^{(1)}
\,.
\label{EOMs_photon_pert}
\end{align}
Thus $A^{(1)}$ is quadratic in $\phi$. 
By plugging the solution $A_i$ into the action \eqref{setup_d}, integrating by parts kinetic terms and using equations \eqref{EOMs_photon_free}, \eqref{EOMs_photon_pert}, at leading order one finds the implicit effective action \eqref{S_eff_implicit} with
\begin{align}
V^{\rm phot}_{\rm eff,quartic} [\phi,\phi^\dag]
=
\frac{1}{2} A_{\mu}^{(1)} [\phi,\phi^\dag] J^\mu [\phi,\phi^\dag] 
\sum_i g_i^2 q_i^2 
\,.
\label{eff_photons}
\end{align}
Notice that by integrating by parts, boundary terms are also produced. We can however neglect them, as we will clarify in a moment.

\medskip
Given the effective interaction \eqref{eff_photons}, \eqref{be_implicit} prescribes to compute $A^{(1)}_\mu [\psi^*_0,\psi_0]  J^\mu [\psi^*_0,\psi_0]$. While the current has a very simple form
\begin{align}
\label{J_explicit}
J^\mu [\psi^*_0,\psi_0] = 2N^2_\Delta(\Delta y^{2\Delta+2},0,0,0,0) 
\,,
\end{align}
the potential $A^{(1)}_\mu [\psi^*_0,\psi_0]$ is found by solving \eqref{EOMs_photon_pert} with source \eqref{J_explicit}. Since $J^{\mu\neq t}=0$, we can adopt the temporal gauge $A^{(1)}_{\mu\neq t}=0$ and, since the source is purely a function of $y$, we look for a solution which is just a function of $y$, $A^{(1)}_t(y)$. The equation of motion becomes (where primes denote derivatives with respect to $y$)
\begin{align}
y \left(\left(3 y^2+1\right)
   A'^{(1)}_t +y \left(y^2-1\right)
   A''^{(1)}_t \right)
   =
   2 \Delta 
   N_{\Delta }^2 y^{2 \Delta }
   \,,
\end{align}
and it is solved by
\begin{align}
A^{(1)}_t = \frac{N^2_\Delta}{2} \frac{y^2 - y^{2\Delta}}{(\Delta-1)(1-y^2)}
\,.
\label{Apert_explicit}
\end{align}
Here the integration constants have been fixed by requiring smoothness at the AdS origin, $y=1$, and vanishing of $A^{(1)}_t$ at the boundary, $y=0$. In fact, the boundary behaviour of $F_i$ in \eqref{EOMs_photons} is fixed, and it must be respected at all orders in perturbations. This means that the perturbation to the field strength $F^{(1)}_{\mu\nu}$ must fall off at least as fast as the free solution, $F^{(0)}_{\mu\nu}\sim y$ at $y\to0$. This is indeed the case for \eqref{Apert_explicit}, $A^{(1)}_{\mu}\sim (y^2,0,0,0,0)$, $F^{(1)}_{yt}\sim y$.

We can now address boundary terms obtained by integration by parts. Among these, only the quartic one is relevant, but it vanishes due to the asymptotic behaviour of the solution: 
\begin{align}
\int \d^5x 
~ \del_\mu 
\left[ 
\sqrt{-g} 
F^{(1) \mu\nu} A^{(1)}_{\nu}
\right]
\propto \lim_{y\to0}
\int_\calb \d^4 x \sqrt{-g} ~ g^{yy} g^{\nu\mu}
F^{(1)}_{y \nu} A^{(1)}_{\mu} 
= 0
\,,
\label{BTs_photon}
\end{align}
where we also used $\sqrt{-g}\sim y^{-5}$, $g^{\mu\nu}\sim y^2$ as $y\to0$. 

\medskip
We can finally substitute \eqref{J_explicit} and \eqref{Apert_explicit} into \eqref{eff_photons} and perform the integral \eqref{be_implicit} to obtain (reinstating $L$)
\begin{align}
\label{be_photon}
\gamma^{\rm phot} = 
\frac{\pi^2 N_{\Delta }^4}{L^2(2 \Delta-1)}  \sum_i g_i^2 q_i^2 
\,.
\end{align}
As for $\gamma^{V}$, the extremum at $y=0$ does not contribute to the integral.

\subsection{Graviton exchange}
\label{sec:graviton}

We can repeat the same procedure for the graviton. First of all, let us consider the action obtained after integrating out the photons 
\begin{align}
S[\phi,h] 
&= 
\int \d^dx \sqrt{-g} 
\bigg[ 
\frac{1}{\kappa^2} \left( \frac{R}{2} + 6 \right) 
- |\del\phi|^2
- m^2|\phi|^2 
- V 
- V^{\rm phot}_{\rm eff}
\bigg]
\;.
\end{align}
We introduce the graviton through $g_{\mu\nu} \to g_{\mu\nu} + \kappa h_{\mu\nu}$. At leading order in $\kappa$ (neglecting sub-leading terms involving other couplings), integrating by parts the kinetic term one obtains
\begin{align}
S[\phi,h] =  
\int \d^5x \sqrt{-g} 
\bigg[ 
- \frac{1}{2} h^{\mu\nu} \Delta^{\rho\sigma}_{\mu\nu} h_{\rho\sigma}
+ \frac{\kappa}{2} h^{\mu\nu} T_{\mu\nu}
- |\del\phi|^2 - m^2|\phi|^2 
\bigg]
\,,
\label{Sgphi}
\end{align}
where 
\begin{align}
\nn
\Delta^{\rho\sigma}_{\mu\nu}h_{\rho\sigma} 
&\equiv 
2 h_{\mu\nu} - g_{\mu\nu} h 
- \frac{1}{4} \nabla_\nu\nabla_\mu h 
+ \frac{1}{4} \nabla_\rho\nabla_\mu h_\nu^{~\rho}
+ \frac{1}{4} \nabla_\rho\nabla_\nu h_\mu^{~\rho} 
\\ \nn
& ~~~~~
- \frac{1}{4} \nabla^2 h_{\mu\nu} 
- \frac{1}{4} g_{\mu\nu} 
(\nabla_\rho \nabla_\sigma h^{\rho\sigma} 
- \nabla^2 h) \,, \\
\label{Tmunu} 
T_{\mu\nu} [\phi,\phi^\dag] 
&= 
g_{\mu\nu} \left(-|\del\phi|^2)-m^2|\phi|^2\right) 
+  \left(\del_\mu\phi^\dag \del_\nu\phi + h.c. \right) 
\,.
\end{align}
The action \eqref{Sgphi} is all we need in order to compute the graviton contribution to $\gamma$ at leading order $\calo(\kappa^2)$. One can indeed show that neglected terms, such as $\calo(\frac{h}{\kappa})$, $\calo(\kappa h^3)$ and higher order terms, do not contribute to 4-point diagrams relevant for $\gamma$. We will expand on boundary terms coming from the integration by parts below.

At $\calo(\kappa)$, the equation of motion is 
\begin{align}
\Delta^{\rho\sigma}_{\mu\nu}h_{\rho\sigma} 
= \frac{\kappa}{2} T_{\mu\nu} 
\,.
\label{EOMs_grav}
\end{align}
We can solve \eqref{EOMs_grav} in the perturbative expansion $h=h^{(0)}+\kappa h^{(1)}$, where free part and perturbation solve respectively 
\begin{align}
\label{EOMs_grav_free}
& \Delta^{\rho\sigma}_{\mu\nu} h^{(0)}_{\rho\sigma} 
= 0
\,, \\
& \Delta^{\rho\sigma}_{\mu\nu} h^{(1)}_{\rho\sigma} 
= \frac{T_{\mu\nu}}{2} 
\,.
\label{EOMs_grav_pert}
\end{align}
Analogously to $A^{(1)}$, also $h^{(1)}$ depends quadratically on $\phi$. By plugging $h=h^{(0)}+\kappa h^{(1)}$ into \eqref{Sgphi} and using \eqref{EOMs_grav_free}, \eqref{EOMs_grav_pert}, at leading order we get an effective action \eqref{S_eff_implicit} 
with quartic contribution
\begin{align}
V^{\rm grav}_{\rm eff,quartic}[\phi,\phi^\dag]  
= - \frac{\kappa^2}{4} h^{(1) \mu\nu}[\phi,\phi^\dag]  T_{\mu\nu}[\phi,\phi^\dag] 
\,.
\label{eff_grav}
\end{align}
We can now explicitly compute \eqref{eff_grav} and its contribution to \eqref{be_implicit}. Similarly to the photon case, we need $h^{(1)}_{\mu\nu} [\psi^*_0,\psi_0] T^{\mu\nu} [\psi^*_0,\psi_0]$. The stress-energy tensor is 
\begin{align}
\label{Tmunu_expl}
T^\mu_{\ \nu} [\psi^*_0,\psi_0] 
=  \Delta N_\Delta^2 y^{2\Delta} \text{diag} \left( 4-2\Delta,4,4-2\Delta+2y^2\Delta,4-2\Delta+2y^2\Delta,4-2\Delta+2y^2\Delta \right)
\,,
\end{align}
where we used the mass--dimension relation $m^2=\Delta(\Delta-4)$. We then solve \eqref{EOMs_grav_pert} by exploiting the symmetry of the source \eqref{Tmunu_expl}, which allows us to take the $h^{(1)}_{\mu\nu}$ Ansatz where $h^{(1)}_{tt}(y),h^{(1)}_{yy}(y)$ are the only non-vanishing components. There are only two independent equations:
\begin{align}
& 3 y^2 \left(y
   \left(y^2-1\right)
   h'^{(1)}_{yy}+2
   \left(y^2+1\right)
   h^{(1)}_{yy} \right)
   =
   4 (\Delta -2) \Delta  N_{\Delta }^2
   y^{2 \Delta }
\,, \\
& 3 y^2 \left(y
   h'^{(1)}_{tt}+2
   h^{(1)}_{tt}+2
   \left(y^2-2\right)
   h^{(1)}_{yy}\right)
   =
   8 \Delta  N_{\Delta }^2 y^{2 \Delta}
\,,
\end{align}
and they are solved by
\begin{align}
h^{(1)}_{tt} 
= \frac{4}{3} \Delta A^{(1)}_t 
= \frac{2}{3} \frac{\Delta N^2_\Delta}{\Delta-1}  \frac{y^2 - y^{2\Delta}}{1-y^2} 
\,, \qquad
h^{(1)}_{yy} 
= \frac{1}{1-y^2} 
\left(
h^{(1)}_{tt} - \frac{2}{3} \Delta N^2_\Delta y^{2\Delta-2}
\right)
\,,
\label{hpert_explicit}
\end{align}
where $A^{(1)}_t$ is given by \eqref{Apert_explicit}. One can check that the solution is smooth at $y=1$ and vanishes at $y=0$ in a way that the correction to the curvature $R^{(1)}$ is sub-leading with respect to the free behaviour $R^{(0)}$. 
We can now comment on boundary terms like those coming from integrating by parts kinetic terms, or the Gibbons-Hawking-York (GHY) term. These are all quadratic in $h$, schematically
\begin{align}
\label{BT_grav}
\lim_{y\to 0} \int_{\calb} \d x^4 \sqrt{-g} g^{-1} g^{-1} g^{-1} h \del h
\sim 
\lim_{y\to 0} \int_{\calb} \d x^4 h^2 
\,.
\end{align} 
The quartic contribution is obtained by taking both $h \to h^{(1)}$, and thus it vanishes. 

\medskip
By substituting \eqref{Tmunu_expl}, \eqref{hpert_explicit} into \eqref{eff_grav} and performing the integral \eqref{be_implicit}, one gets (reinstating $L$): 
\begin{align}
\label{be_graviton}
&\gamma^{\rm grav} 
=
-\frac{2 \pi ^2 (\Delta -2) \Delta
   ^2 \kappa^2 N_{\Delta }^4}{3
   (\Delta -1) (2 \Delta -1) L^4} 
\,.
\end{align}
Again the boundary does not contribute to this integral.

	\subsection{Scalar exchange}
	\label{sec:scalars}
	
We will now compute the contribution to the binding energy given by $\chi$ in \eqref{setup_d} following the same procedure. The generalization to multiple scalars $\chi_i$ with a diagonal mass matrix is straightforward. The equation of motion is
\begin{align}
\label{EOM_scalar}
\square \chi - M^2 \chi = Y |\phi|^2  + \delta |\del\phi|^2
\,,
\end{align}
where $\square \chi \equiv \tfrac{1}{\sqrt{-g}} \del_\mu ( \sqrt{-g} \del^\mu \chi  )$. 
We can integrate out $\chi$ classically using $\chi=\chi^{(0)}+\chi^{(1)}$, where $\chi^{(0)},\chi^{(1)}$ solve respectively
\begin{align}
\label{EOM_scalar_0}
& \square \chi^{(0)} - M^2 \chi^{(0)} = 0
\,, \\
\label{EOM_scalar_1}
& \square \chi^{(1)} - M^2 \chi^{(1)} = Y |\phi|^2 + \delta |\del\phi|^2
\,.
\end{align}
Thus also $\chi^{(1)}$ is quadratic in $\phi$. By plugging $\chi=\chi^{(0)}+\chi^{(1)}$ into \eqref{setup_d}, integrating by parts the kinetic term and using \eqref{EOM_scalar_0}, \eqref{EOM_scalar_1}, one finds an effective potential with leading order quartic terms 
\begin{align}
V^{\rm scal}_{\rm eff,quartic} [\phi,\phi^\dag]
=
\frac{1}{2} \chi^{(1)}[\phi,\phi^\dag] \left( Y |\phi|^2 + \delta |\del\phi|^2 \right) 
\,,
\label{eff_scalar}
\end{align}
up to the usual boundary term produced by integration by parts, which we discuss in a moment. Since $\chi^{(1)}$ is linear in $Y$ and $\delta$, $V^{\rm scal}_{\rm eff,quartic}$ is quadratic in couplings $Y,\delta$.

\medskip
 Analogously to the previous Section, we must solve \eqref{EOM_scalar_1} with $\phi\to\psi^*_0$, and therefore we can make the simple Ansatz $\chi^{(1)}(y)$, and solve 
\begin{align}
\label{chi1_eq} 
N^2_\Delta y^{2\Delta} \left(Y + \delta \Delta^2(1-2y^2)\right) + M^2 \chi_{(1)} + y ( (y^2+3) \chi'_{(1)} + y (y^2-1) \chi''_{(1)} ) = 0
\,.
\end{align}
Here primes denote derivatives with respect to $y$.
Equation \eqref{chi1_eq} can be solved analytically for $M^2=0$, as we will show in Section \ref{sec:massless}, and for $M^2=-4, \delta=0$ as shown in Section \ref{sec:BF}. Moreover, in Section \ref{sec:numerics}, we present numerical results for $\delta=0$ and different values of $M$.

\medskip
It is worth noting that the \emph{asymptotic} behaviour of the solution can be found analytically and in full generality for $M^2>-4$:
\begin{align}
\nn
\chi_{(1)}
~
\stackrel{y\to0}{\longrightarrow} 
~
& y^{2\Delta} \left( \frac{(Y+\delta\Delta^2)N_\Delta^2}{4\Delta(\Delta-2)-M^2} + \calo(y^2) \right) +
\\
&~~ + C_1 y^{2-\sqrt{4+M^2}} (1+ \calo(y^2))
+ C_2 y^{2+\sqrt{4+M^2}} (1+ \calo(y^2))
\,,
\label{asympt}
\end{align}
where $\Delta \neq 1\pm\sqrt{1+\tfrac{M^2}{4}}$. In case the BF bound is saturated, $M^2=-4$, this asymptotic expansion misses a log term, see Section \ref{sec:BF}.
The first term in \eqref{asympt} corresponds to the particular solution while the second and third ones are the homogeneous solution. We must ensure that the asymptotic behaviour \eqref{asympt} does not spoil the free one, namely $\chi_{(1)}$ must fall off at least as $\chi_{(0)}$ towards $y\to0$. 
As explained in Appendix \ref{app:free}, the free solution $\chi_{(0)}$ falls off with $ y^{2+\sqrt{4+M^2}}$ if $\sqrt{4+M^2}\ge1$, while both $y^{2+\sqrt{4+M^2}}$ and $y^{2-\sqrt{4+M^2}}$ fall-offs are possible if $0<\sqrt{4+M^2}<1$. 
This means that if $\chi_{(0)}$ decays with $y^{2+\sqrt{4+M^2}}$, the first and second term in \eqref{asympt} must be respectively sub-leading and absent, namely
\begin{align}
\label{quantization_1}
\Delta > 1+\sqrt{1+\tfrac{M^2}{4}}
\quad
\text{and}
\quad
C_1=0 
\,.
\end{align}
Interestingly, the condition on $\Delta$ is stronger that the unitarity bound $\Delta>1$ and moreover it is stronger the heavier the $\chi$ is. In particular, this means that our perturbative analysis cannot cover the parameter space $M\gg 2\Delta$. In such a case, one may integrate out $\chi$ to obtain contact terms of the types contained in $V$ and its contribution to the binding energy will be contained in $\gamma^V$.
On the contrary, if the free solution falls off with $y^{2-\sqrt{4+M^2}}$, we do not need to impose any constraint on \eqref{asympt}. 

\medskip
We can now comment on the boundary terms coming from integration by parts. These contain the following quartic contribution
\begin{align} 
\lim_{y\to0}
 \int_\calb \d^4 x \sqrt{-g} ~ g^{yy}
 \chi^{(1)}  \del_y  \chi^{(1)}
\sim 
\lim_{y\to0}
\int_\calb \d^4 x ~ y^{-3}
 \chi^{(1)}  \del_y  \chi^{(1)}
\,.
\label{BTs_scal}
\end{align}
This term vanishes if $\chi^{(1)}$ vanishes faster than $y^2$. While this is always the case for $\sqrt{4+M^2}\ge1$ (as discussed above), it seems to suggest that also for $0<\sqrt{4+M^2}<1$ we have to consider the subset of $\chi_{(1)}$'s with $C_1=0$ asymptotically. 
This sounds to be unreasonably restrictive. The situation is even more cumbersome in case the BF bound is saturated. There, $\chi_{(1)}\sim y^2$ asymptotically, and the boundary term \eqref{BTs_scal} cannot vanish. These problems are avoided adopting the Klebanov--Witten action \cite{Klebanov:1999tb}, which formally corresponds to integrating out the kinetic terms 
$|\del\phi|^2$ in \eqref{setup_d} and throwing away the boundary term. The boundary term \eqref{BTs_scal} is then absent. We can simply assume to start with such a scalar action instead of \eqref{setup_d}, and compute the contribution to the binding energy only given by $V^{\rm scal}_{\rm eff,quartic}$ \eqref{eff_scalar}.

	\subsubsection{Massless case}
	\label{sec:massless}

For $M=0$, we can solve \eqref{chi1_eq} analytically. The two integration constants can be fixed by requiring that the solution is smooth at $y=1$ and vanishes at $y=0$, to obtain:
\begin{align}
\label{massless_sol}
\chi_{(1)} &=
\frac{N_{\Delta}^2 }{4 \Delta (\Delta -2) }
\left[  
\frac{\Delta(Y-\delta \Delta(\Delta-4))}{\Delta -1}
\left(
\frac{y^2}{y^2-1}
- \log (1-y^2)
+ f_\Delta (y)
\right)
+ y^{2\Delta}(Y+\delta\Delta^2)
\right]
 \,,
\end{align}
where the first two terms inside the (curved) parenthesis correspond to the homogeneous solution, and the remaining terms correspond to the particular solution, with
\begin{gather}
\label{partic_sol}
f_\Delta (y) 
= 
B_{y^2}(\Delta+1,-1)
+ (\Delta -1) B_{y^2}(\Delta +1,0)
\,, \\
\nn
B_{x}(a,b)
\equiv 
\int_0^x \d t \, t^{a-1} (1-t)^{b-1}
\,,
\end{gather}
where $B_{x}(a,b)$ are incomplete Beta functions. 
As expected, the homogeneous part in \eqref{massless_sol}, falls-off as the free part, with $y^4$, and the particular solution falls-off with $y^{2\Delta}$. In order for perturbation theory to make sense, we need $\Delta\ge2$ for $\phi$.\footnote{One can indeed show that also the solution with $\Delta=2$ is perfectly regular.}

\medskip
Plugging the solution \eqref{massless_sol} into \eqref{eff_scalar} and \eqref{be_implicit}, and carefully integrating on $y$, one finds that only the the interior of AdS $y=1$ contributes, obtaining\footnote{In particular, given any function $g(y)$ with primitive $G(y)$, we integrate by parts all terms like
$$
\int \d y ~  f_\Delta (y) g(y)
=
 f_\Delta (y) G(y)
- \int \d y ~  f_\Delta' (y) G(y)
\,.
$$}
\begin{align}
\nn
\gamma^{\rm scal}
& = 
\frac{\pi^2 N_{\Delta }^4 }{8 (\Delta-2)^2 (\Delta-1)^2}
\bigg[
Y^2
\left(
1-\Delta+\frac{1}{\Delta}+\frac{4}{\Delta-1}+\frac{2}{2\Delta-1}
+4 H_{\Delta-2} - 2 H_{2 \Delta}
\right) 
\\ \nn
& ~~
+ \frac{2 Y\delta}{L^2(2\Delta-1)}
\bigg(
4 + 7\Delta (\Delta-1) - 9\Delta^3 + 2 \Delta^4
+2 \Delta (2\Delta^2-9\Delta+4)
\left(H_{2 \Delta}-2 H_{\Delta}\right)
\bigg) 
\\
& ~~
+ \frac{\delta^2 \Delta^2}{L^4} 
\left(
-6+\frac{2}{2\Delta-1}
-\Delta(\Delta(\Delta-7)+11)
-2 (\Delta-4)^2
\left(H_{2 \Delta}-2 H_{\Delta}\right)
\right) 
\bigg]
\,.
\label{be_scalar_massless}
\end{align}
Here $H_n$ is the harmonic number $H_n\equiv \sum_{k=1}^n \frac{1}{k}$, and we reintroduced $L$.

	\subsubsection{The case $M^2=-4$, $\delta=0$}
	\label{sec:BF}
	
In case $\delta=0$ and the BF bound is saturated, $M^2=-4$, equation \eqref{chi1_eq} can be solved analytically:
\begin{align}
\chi_{(1)} &=
\frac{Y N_{\Delta}^2}{4 (\Delta -1)^2 }
\left(
\frac{-y^{2\Delta}-4y^2(\Delta-1)^2(c_1+c_2 \log (y))}{y^2-1}
\right)
\,.
\label{BF_naive}
\end{align}
The first term in the brackets corresponds to the particular solution, and the rest is the homogenous solution. We can easily fix $c_1$ requiring smoothness at $y=1$, obtaining $c_1=-\tfrac{1}{4(\Delta-1)^2}$. On the other hand, $c_2$ must vanish in order for the perturbative expansion around the free solution $\chi_{(0)}$ to make sense. Since $\chi_{(0)}$ falls-off as $y^2$ (see Appendix \ref{app:free}), if $c_2\neq0$ then there exist a region close to the boundary where  $\chi_{(1)}\ge\chi_{(0)}$, no matter how small $Y$. In other words, $c_2=0$ in order to respect the boundary conditions chosen for $\chi$ \cite{Breitenlohner:1982bm,Breitenlohner:1982jf}.\footnote{We thank Ofer Aharony for a clarification on this point.}
The solution is therefore
\begin{align}
\label{BF_sol}
\chi_{(1)} &=
\frac{Y N_{\Delta}^2 }{4 (\Delta -1)^2 }
\left(
\frac{-y^{2\Delta}+y^2}{y^2-1}
\right)
\,.
\end{align}
The unitarity bound $\Delta>1$ then ensures that $\chi_{(1)}$ falls-off at least as $\chi_{(0)}$.

\medskip
Again, in computing the binding energy \eqref{be_implicit}, one finds that only the interior of AdS yields a non-trivial contribution, which is 
\begin{align}
\label{be_BF}
\gamma^{\rm scal}
= 
- \frac{Y^2 \pi^2 N_{\Delta }^4 }{8 (\Delta-1)^3}
\,.
\end{align}

	\subsubsection{Generic $M$, with $\delta=0$}
	\label{sec:numerics}

\begin{figure}[ht!]
\begin{center}
\hspace{-15pt}
\includegraphics[width=0.5\textwidth]{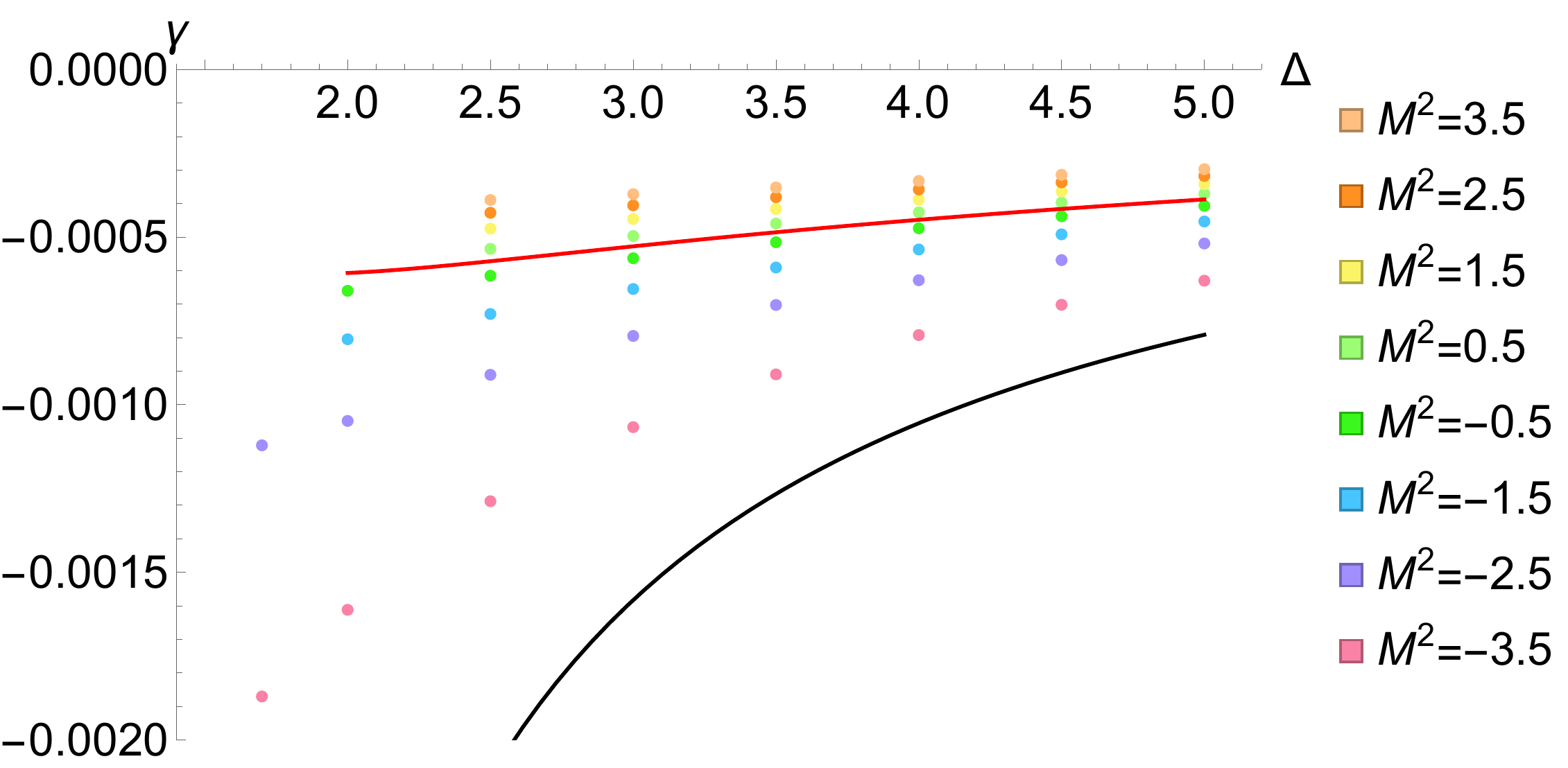}
\hspace{0pt}
\includegraphics[width=0.5\textwidth]{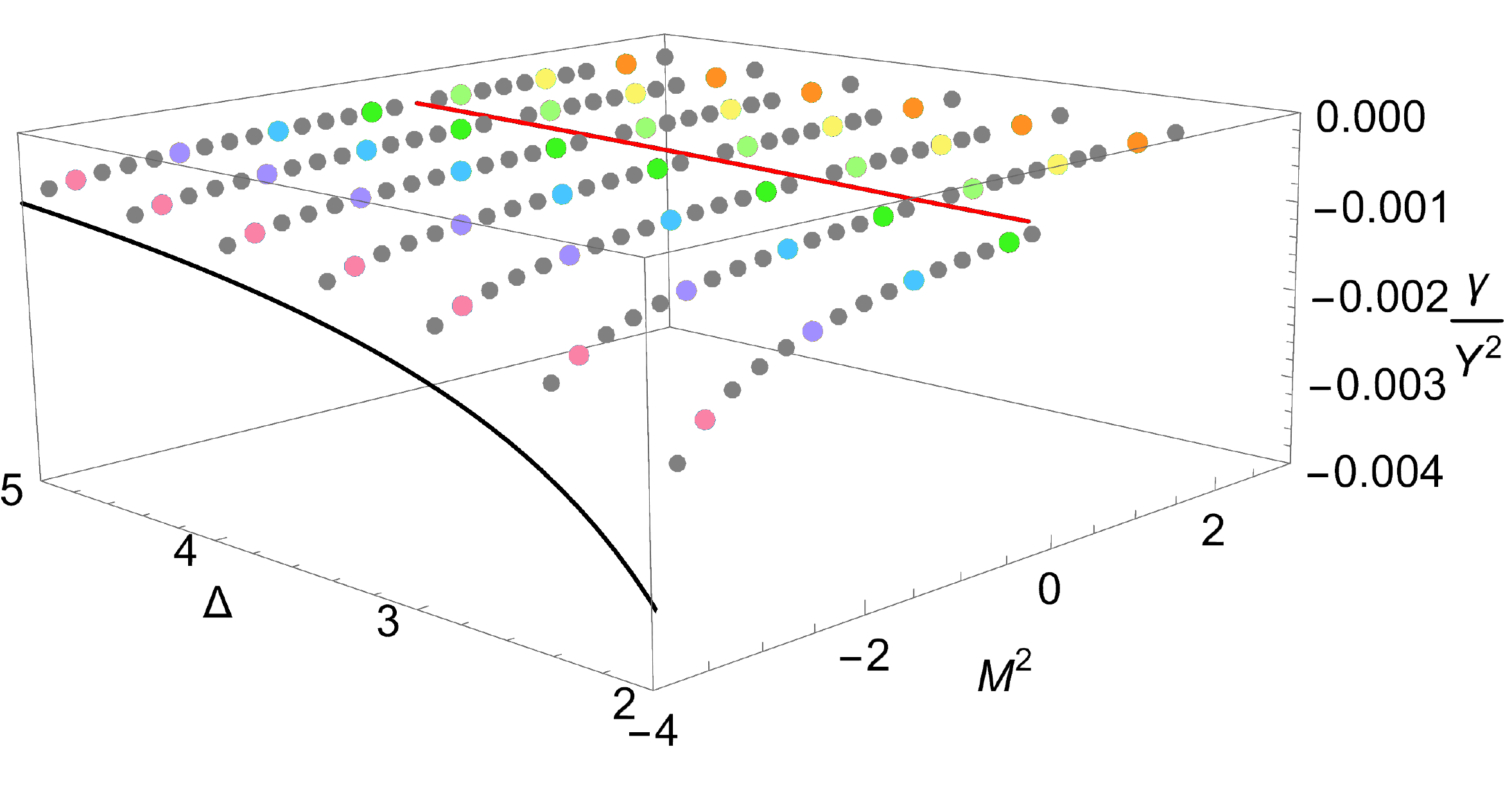}
\caption{\label{fig:numerics}\textit{Contribution to the binding energy $\gamma$ for different scaling dimensions of $\phi$, $\Delta$, and scalar mass of $\chi$, $M$. Black and red lines represent binding energies for the analytically accessible cases $M^2=-4$ \eqref{be_BF} and $M^2 = 0$ \eqref{be_scalar_massless} respectively.}}
\end{center}
\end{figure}

In order to compute the binding energy in general, we need to rely on numerical methods. Here, we do so by setting $\delta=0$ for simplicity, so that we can compute the numerical values of  $\tfrac{\gamma(\Delta)}{Y^2}$ for different values of the mass parameter $M$. We solve equation \eqref{chi1_eq} with a shooting method from $y=1$ to the boundary $y \rightarrow 0$. Requiring smoothness at $y=1$ fixes $\chi'(1)$ as a function of $\chi(1)$, and the last degree of freedom that identifies the solution (i.e., the right value of $\chi_{(1)}$ at $y=1$) is set by requiring the match to the asymptotic behaviour \eqref{asympt} close to $y=0$. As discussed above, for $M^2\ge-3$, the asymptotic must satisfy condition \eqref{quantization_1}, whereas for $-4<M^2<-3$ two possible fall-offs are allowed. For the sake of convenience, we focus on the branch of solutions satisfying \eqref{quantization_1} also in this mass range.
\footnote{More precisely, the match of the numerical solution $\chi_{(1)}$ to the asymptotic $\chi_{(1)}^{\rm asympt}$ provided in \eqref{asympt} is obtained by requiring that, at small $y$,
\begin{align*}
\left( \frac{\chi_{(1)}}{y^{2+\sqrt{4+M^2}}} \right)' 
\overset{!}{=}
\left( \frac{\chi_{(1)}^{\rm asympt}}{y^{2+\sqrt{4+M^2}}} \right)' \bigg|_{C_1=0}
=
\frac{Y N_\Delta^2(2\Delta-2-\sqrt{4+M^2})}{4\Delta(\Delta-2)-M^2} 
y^{2\Delta-3-\sqrt{4+M^2}} 
\,.
\end{align*}
}
We can thus fix $\Delta$ and $M$, find $\chi_{(1)}$, and plug it into \eqref{eff_scalar}, \eqref{be_implicit} to compute the binding energy. Results are collected in Table \ref{tab:numerics} and plotted in Fig.~\ref{fig:numerics}. We have highlighted analytic results obtained for $M^2=(0,-4)$, see \eqref{be_scalar_massless} and \eqref{be_BF} respectively.
\begin{table}[t!]
	\scalebox{0.75}{
\begin{tabular}{S|rSSSSSSSSS} \toprule
	{$\gamma/Y^2 \times 10^{4}$} & {$\Delta = 1.3$} & {$\Delta = 1.5$} &{$\Delta = 1.7$} &{$\Delta = 2$} & {$\Delta = 2.5$} & {$\Delta = 3$} & {$\Delta = 3.5$} & {$\Delta = 4$} & {$\Delta = 4.5$} & {$\Delta = 5$} \\ \midrule
	\rowcolor{Gray}{$M^2=-4$} & -106 & -63.3 & -45.2 & -31.6 & -21.1 & -15.8 & -12.6 & -10.5 & -9.05 & -7.92 \\ \midrule
	{$M^2=-3.75$}& -30.5&-27.1&-23.6 & -19.5 & -14.9 & -12.0 & -10.1 & -8.68 & -7.61 & -6.79  \\
	{$M^2=-3.5$} && -20.5& -18.7& -16.1 & -12.9 & -10.70 & -9.10 & -7.93 & -7.02 & -6.30  \\
	{$M^2=-3$} &&& -13.8 & -12.6&-10.6&-9.02&-7.86&-6.95&-6.24&-5.65 \\
	{$M^2=-2.5$}&&&  & -10.5&-9.11&-7.95&-7.03&-6.29&-5.69&-5.19     \\ 
	{$M^2=-2$} &&& & -9.08&-8.08&-7.16&-6.40&-5.78&-5.26&-4.83 \\
	{$M^2=-1.5$} &&& & -8.05&-7.30&-6.55&-5.91&-5.37&-4.92&-4.54  \\
	{$M^2=-1$} &&& &  -7.25&-6.67&-6.05&-5.50&-5.03&-4.63&-4.29  \\
	{$M^2=-0.5$}&&&  &-6.60&-6.15&-5.63&-5.15&-4.74&-4.38&-4.07   \\ \midrule
	\rowcolor{Gray}	{$M^2=0$}&&& & -6.07 & -5.72 & -5.28 & -4.86 & -4.49 & -4.16 & -3.88 \\ \midrule
	{$M^2=0.5$} &&& & & -5.35&-4.97&-4.60&-4.26&-3.97&-3.71  \\
	{$M^2=1$} &&& & & -5.03&-4.70&-4.37&-4.07&-3.80&-3.56  \\
	{$M^2=1.5$} &&& && -4.75&-4.46&-4.16&-3.89&-3.64&-3.42  \\
	{$M^2=2$} &&& & & -4.50&-4.25&-3.98&-3.73&-3.50&-3.29 \\
	{$M^2=2.5$} &&& & & -4.28&-4.05&-3.81&-3.58&-3.37&-3.18  \\
	{$M^2=3$} &&& && -4.08&-3.88&-3.66&-3.45&-3.25&-3.07 \\
	{$M^2=3.5$} &&& && -3.89&-3.72&-3.52&-3.33&-3.14&-2.97 \\ \bottomrule
\end{tabular}
}
\caption{\label{tab:numerics}\textit{Binding energy contributions from scalar exchanges of different mass $M$. These results correspond to solutions $\chi_{(1)}$ satisfying \eqref{quantization_1}. Missing values do not satisfy this condition. We have highlighted rows corresponding to analytic results, obtained for $M^2=(0,-4)$. }}
\end{table}

\section{Binding energy in $\ads_4$}
\label{sec:4d}

We can repeat the same procedure for an effective theory \eqref{setup_d} in $d=4$, with perturbation parameters 
\begin{align}
\label{small_int}
g_i\ll 1
\,, \quad
\frac{\kappa}{L}\ll 1
\,, \quad
|a|\ll 1
\,, \quad
\frac{|b|}{L^2}\ll 1
\,, \quad
\frac{|c|}{L^2}\ll 1
\,, \quad
|Y| L\ll 1
\,, \quad
\frac{|\delta|}{L}\ll 1
\,.
\end{align}
After integrating out classically photons, graviton and $\chi$, one finds the same (implicit) effective action as in $d=5$, namely:
\begin{align}
S[\phi] =  
\int \d^4x \sqrt{-g} 
\bigg[ 
- |\del\phi|^2 - m^2|\phi|^2 
- V 
- V^{\rm phot}_{\rm eff}
- V^{\rm grav}_{\rm eff}
- V^{\rm scal}_{\rm eff}
\bigg]
\,, 
\label{Seff_4d}
\end{align}
where $V$ is the contact term \eqref{V_quartic} and again
\begin{align}
\label{eff_photons_4}
& V^{\rm phot}_{\rm eff,quartic} [\phi,\phi^\dag] 
= \frac{1}{2} A_{\mu}^{(1)} [\phi,\phi^\dag] J^\mu [\phi,\phi^\dag] 
\sum_i g_i^2 q_i^2
\,, \\
\label{eff_grav_4}
& V^{\rm grav}_{\rm eff,quartic}[\phi,\phi^\dag]  
= - \frac{\kappa^2}{4} h^{(1) \mu\nu}[\phi,\phi^\dag]  T_{\mu\nu}[\phi,\phi^\dag] 
\,,\\
\label{eff_scalar_4}
& V^{\rm scal}_{\rm eff,quartic} [\phi,\phi^\dag]
=\frac{1}{2} \chi^{(1)}[\phi,\phi^\dag] \left( Y |\phi|^2 + \delta |\del\phi|^2 \right) 
\,.
\end{align}
Perturbations $A^{(1)},h^{(1)},\chi^{(1)}$ obey equations \eqref{EOMs_photon_pert},\eqref{EOMs_grav_pert} and \eqref{EOM_scalar_1} in $d=4$, and we will solve them explicitly in the next Sections. There, we will also comment on boundary terms coming from integration by parts.

\subsection{Contribution from contact terms}
\label{sec:V_4}

Proceeding in the same order as in Section \ref{sec:5d}, let us first compute the contribution to the binding energy \eqref{be_implicit} given by $V[\psi_0^*,\psi_0]$. The lowest energy mode is now (see \eqref{app:psi0_4})
\begin{align}
\psi^*_0 =  N_\Delta e^{it\Delta} y^\Delta
\,, \quad 
N_\Delta =  \frac{1}{\pi^{3/4}} \sqrt{\frac{\Gamma(\Delta+1)}{2\Delta\Gamma(\Delta-1/2)}}
\,, \quad
m^2 = \Delta(\Delta-3)
\,,
\end{align}
and $\Delta>\frac{1}{2}$ is the unitarity bound. 
One finds that the integral converges at $y\to0$ as long as $\Delta>\frac{3}{4}$, and it gives
\begin{align}
\label{be_quartic_4}
&\gamma^{V} = 
\pi^{3/2} N_{\Delta}^4 
\left(2a-\frac{b}{L^2} (2\Delta - 3) \Delta + 4\Delta^2 \frac{c}{L^2} \right)
\frac{\Gamma\left(2\Delta-\tfrac{3}{2}\right)}{\Gamma(2\Delta) L} 
\,.
\end{align}

\subsection{Photon exchanges}
\label{sec:photons_4}

Focusing on the $i$-th photon, we solve \eqref{EOMs_photon_pert} noticing that the source is 
\begin{align}
\label{J_explicit_4}
J^\mu [\psi_0^*,\psi_0] = 2N^2_\Delta (\Delta y^{2\Delta+2},0,0,0) 
\,.
\end{align}
Choosing the temporal Ansatz and requiring smoothness at $y=1$ and vanishing at $y=0$, we find:
\begin{align}
A^{(1)}_t = 
\frac{N_\Delta^2}{2} \frac{y }{\sqrt{1-y^2}} \left( 2 \pi^{1/2} \frac{\Gamma(\Delta+1/2)}{(2\Delta-1)\Gamma(\Delta)} - B_{y^2}(\Delta-1/2,1/2) \right) 
\,,
\label{Apert_explicit_4}
\end{align}
where $B_y^2$ is the incomplete Beta function defined in \eqref{partic_sol}. 
Asymptotically, the perturbation behaves as $A^{(1)}_{t}\sim y$, and thus $F^{(1)}_{yt}\sim \calo(1)$ does not spoil the free solution, $F^{(0)}_{\mu\nu}\sim \calo(1)$. Furthermore, one can check that the boundary term analog to \eqref{BTs_photon} vanishes.
The binding energy \eqref{be_implicit} given by all photons in $V^{\rm phot}_{\rm eff,quartic}$ \eqref{eff_photons_4} is therefore:
\begin{align}
\label{be_photon_4}
\gamma^{\rm phot} = 
2 \pi^{3/2} \frac{N_{\Delta }^4}{L} 
\frac{\Gamma(2\Delta-1/2)}{\Gamma(2\Delta)}  \sum_i g_i^2 q_i^2 
\,,
\end{align}
where the boundary does not contribute to this integral.

\subsection{Graviton exchange}
\label{sec:graviton_4}

The stress-energy tensor is 
\begin{align}
T^\mu_{\ \nu} [\psi_0^*,\psi_0] 
= \Delta N_\Delta^2 y^{2\Delta} \text{diag} \left( 
3-2\Delta,3,3-2\Delta+2y^2\Delta,3-2\Delta+2y^2\Delta
\right)
\,,
\end{align}
where we used $m^2=\Delta(\Delta-3)$. We thus solve \eqref{EOMs_grav_pert} using the same Ansatz as in Section \ref{sec:graviton} to find:
\begin{align}
h^{(1)}_{tt} = \Delta A_t^{(1)}
\,, \qquad
h^{(1)}_{yy} = 
\frac{1}{1-y^2} 
\left(
h^{(1)}_{tt} - \Delta N^2_\Delta y^{2\Delta-2}
\right)
\,,
\label{hpert_explicit_4}
\end{align}
where $A_t^{(1)}$ is \eqref{Apert_explicit_4}. Once again, one can check that the correction to the curvature $R^{(1)}$ does not spoil the leading behaviour $R^{(0)}$ asymptotically, and that the boundary term analog to \eqref{BT_grav} vanishes. The integral for the binding energy \eqref{be_implicit} converges if $\Delta>\frac{3}{4}$, yielding:
\begin{align}
\label{be_graviton_4}
\gamma^{\rm grav} = 
- \pi^{3/2} \kappa^2 N^4_\Delta \Delta^2 (2\Delta-3) 
\frac{\Gamma(2\Delta-3/2)}{\Gamma(2\Delta) L^3} 
\,.
\end{align}

	\subsection{Massless scalar $\chi$ exchange}
	\label{sec:scalars_4}
	
Analogously to Section \ref{sec:scalars}, the equation for $\chi^{(1)}$ is
\begin{align}
\label{chi1_eq_4} 
N^2_\Delta y^{2\Delta} \left(Y + \delta \Delta^2(1-2y^2)\right) + M^2 \chi_{(1)} + y ( (y^2+2) \chi'_{(1)} + y (y^2-1) \chi''_{(1)} ) = 0
\,,
\end{align}
and we will solve it analytically in the massless case. 

Setting $M=0$ and fixing the integration constants in the usual way, the solution to \eqref{chi1_eq_4} is 
\begin{align}
\nn
\chi_{(1)} 
&= 
\frac{\sqrt{\pi} N_\Delta^2 (Y-\delta (\Delta-3)\Delta)}{4} \frac{\Gamma(\Delta-3/2)}{\Gamma(\Delta)} \left( \arcsin(y) - \frac{y}{\sqrt{1-y^2}} \right) 
\\ \nn
& ~~~~
+ \frac{N_\Delta^2(Y+\delta \Delta^2)}{2} \frac{y^{2\Delta} \,_3 F_2(1,\Delta,\Delta;\Delta-1/2;\Delta+1;y^2)}{\Delta(2\Delta - 3)}
\\
& ~~~~
- N_\Delta^2 \delta \Delta^2 
 \frac{y^{2\Delta+2} \,_3 F_2(1,\Delta+1,\Delta+1;\Delta+1/2;\Delta+2;y^2)}{(\Delta + 1)(2\Delta - 1)}
 \,.
 \label{massless_sol_4}
\end{align}
As expected, the homogeneous part (first line) of the solution falls-off as the free part, with $y^3$. Thus, the particular part (second and third line) must fall fall-off with $\Delta\ge\frac{3}{2}$.\footnote{Also the solution with $\Delta=\frac{3}{2}$ is perfectly regular.} Plugging the solution \eqref{massless_sol_4} into \eqref{be_implicit} and integrating, one finds for $\Delta>\frac{3}{2}$ 
\begin{align}
\label{be_scalar_massless_4}
\gamma^{\rm scal} = Y^2 L \gamma_1 + \frac{\delta^2}{L^3} \gamma_2 + Y \frac{\delta}{L} \gamma_3
\,, 
\end{align}
\begin{align}
\nn & 
\gamma_1 
= \frac{\pi^{3/2} N_\Delta^4}{8} 
\bigg( 
\frac{\pi^{3/2} \Gamma^2(\Delta-\sfrac{3}{2}) }{\Gamma^2(\Delta)}
- \frac{4 \Gamma(\Delta-\sfrac{3}{2})}{(\Delta-1)^2 \Gamma(\Delta-1)}
- \frac{8 \sqrt{\pi} \calf_1}{(\Delta-1)(2\Delta-3)^2}
+ \frac{8 \Gamma(2\Delta-\sfrac{3}{2})  \calf_2}{(2\Delta-3)\Gamma(2\Delta+1)}
\bigg) 
\\ \nn 
& 
\gamma_2
= \frac{\pi^{3/2} N_\Delta^4 \Delta^2}{8} 
\bigg( 
\frac{\pi^{3/2} (\Delta-3)^2 \Gamma^2(\Delta-\sfrac{3}{2} ) }{\Gamma^2(\Delta)}
- \frac{4 \Gamma(\Delta-\sfrac{3}{2})}{(\Delta-1)^2 \Gamma(\Delta-3)}
+ \frac{8 \sqrt{\pi} \Delta (\Delta-3) \calf_1}{(2\Delta-3)^2(\Delta-1)}
\\ \nn
& ~~~~~~~
- \frac{32 \sqrt{\pi} \Delta (\Delta-3) \calf_3}{(2\Delta-3)(2\Delta-1)^2}
- \frac{2^{5-4\Delta} \sqrt{\pi} \Delta \Gamma(4\Delta-2) \calf_5}{(\Delta+1)\Gamma^2(2\Delta)}
+ \frac{4 \Delta \Gamma(2\Delta-\sfrac{3}{2})(2 \Delta \calf_2 - (4\Delta-3)\calf_4)}{(2\Delta-3)\Gamma(2\Delta+1)}
\\ \nn
& ~~~~~~~
+ \frac{32 \Delta^2 \Gamma(2\Delta+\sfrac{1}{2}) \calf_6}{(2\Delta-1) \Gamma(2\Delta+3)}
\bigg) 
\\ \nn 
& 
\gamma_3
= \frac{\pi^{3/2} N_\Delta^4 }{4} 
\bigg( 
\frac{ 2\Delta(2\Delta-5) \Gamma(\Delta-\sfrac{3}{2})}{(\Delta-1)^2\Gamma(\Delta-1)}
- \frac{\pi^{3/2} (\Delta-3)\Delta \Gamma^2(\Delta-\sfrac{3}{2}) }{\Gamma^2(\Delta)}
- \frac{12 \sqrt{\pi} \Delta \calf_1}{(2\Delta-3)^2(\Delta-1)}
\\ \nn
& ~~~~~~~
+ \frac{16 \sqrt{\pi} \Delta^2 \calf_3}{(2\Delta-3)(2\Delta-1)^2}
+ \frac{\Gamma(2\Delta-\sfrac{3}{2})(4\Delta \calf_2 - (4\Delta-3)\calf_4)}{(2\Delta-3)\Gamma(2\Delta)}
- \frac{4^{2-2\Delta} \sqrt{\pi} \Delta \Gamma(4\Delta-2) \calf_5}{(\Delta+1)\Gamma^2(2\Delta)}
\bigg) 
\,,
\end{align}
where $\calf_i$ are the following hypergeometric functions:
\begin{align}
\nn
\calf_1 &= 
\,_3 F_2 \left( -\tfrac{1}{2},\Delta-\tfrac{3}{2},\Delta-1;\Delta-\tfrac{1}{2},\Delta-\tfrac{1}{2};1 \right)
\,, \\
\nn
\calf_2 &=  \,_4 F_3 \left( 1,\Delta,\Delta,2\Delta-\tfrac{3}{2};\Delta-\tfrac{1}{2},2\Delta,\Delta+1;1 \right)
\,, \\
\nn
\calf_3 &=  \,_3 F_2 \left( -\tfrac{1}{2},\Delta-\tfrac{1}{2},\Delta;\Delta+\tfrac{1}{2},\Delta+\tfrac{1}{2};1 \right)
\,, \\
\nn
\calf_4 &= \,_4 F_3 \left( 1,\Delta,\Delta,2\Delta-\tfrac{1}{2};\Delta-\tfrac{1}{2},\Delta+1,2\Delta+1;1 \right)
\,, \\
\nn
\calf_5 &=  \,_4 F_3 \left( 1,\Delta+1,\Delta+1,2\Delta-\tfrac{1}{2};\Delta+\tfrac{1}{2},\Delta+2,2\Delta+1;1 \right)
\,, \\
\calf_6 &= \,_4 F_3 \left( 1,\Delta+1,\Delta+1,2\Delta+\tfrac{1}{2};\Delta+\tfrac{1}{2},\Delta+2,2\Delta+2;1 \right)
\,.
\label{hypergeometrics}
\end{align}

	\section{Tests in special cases}
	\label{sec:tests}
	
	In this Section we will test our expressions for the binding energy by studying special cases where we should reach expected results. The first is the flat space limit, where we expect to recover the known flat space results. The second is when $\phi$ is a charged BPS state in a supersymmetric vacuum, in which case the self-binding energy should vanish exactly.
	
	\subsection{Flat space limit}
	\label{sec:flat_limit}
	
It is meaningful to explore the flat spacetime limit of the total binding energy. We expect that requiring positive self-energy, $\gamma\ge0$, reproduces the WGC with scalar fields bounds, both in $d=4,5$ \cite{Palti:2017elp,Lee:2018spm}.\footnote{Notice that massless scalars are dimensionless in \cite{Lee:2018spm}, thus an overall factor $M^{2-d}_d$ (for any dimension $d$) appears in front of the scalar contribution there.} 
This limit is achieved by taking $L\to\infty$ while taking all the other effective parameters, couplings and masses fixed. This means that we have to take $\Delta\to\infty$, and analogously $\Delta_\chi\to\infty$ in the case $M^2=L^{-2}\Delta_\chi(\Delta_\chi-d+1)\neq0$ \cite{Fitzpatrick:2010zm}. 
Since conformal dimensions are real, it should be clear that such a limit only makes sense for $m^2>0,M^2\ge0$. 
For $M=0$ we expect to recover the standard formulation of the WGC with scalar fields \cite{Palti:2017elp}, and so we restrict to this case.  
In order to make contact with the existing literature in flat spacetime, we use the parameterisation $Y=2m\mu$.

\medskip
Let us start with $d=5$. In this case, we can consider the leading order term for \emph{each} contribution to the binding energy, \eqref{be_quartic},\eqref{be_photon},\eqref{be_graviton} and \eqref{be_scalar_massless}, obtaining:
\begin{align}
\label{SWGC_5}
\sum_i q_i^2 g_i^2 - \frac{2}{3} m^2\kappa^2 -  \bigg(  \mu - \frac{m\delta}{2} \bigg)^2 
+ \frac{1}{L} \left( \frac{a}{m} - m(b-2c) \right)
\ge 0
\,,
\end{align} 
where we used that $H_n \sim \log n + \gamma + \calo(1/n)$ for $n\to\infty$, with $\gamma$ the Euler--Mascheroni constant. As expected, in the flat limit quartic terms in $V$ are suppressed by $\calo(1/L)$. In the case $\delta=0$, we precisely reproduce the expressions in \cite{Palti:2017elp,Lee:2018spm}. The case with $\delta \neq 0$ shows that there is an additional term which should be accounted for. Dimensional analysis tells us that the contribution from $\delta$ 
(with mass dimension $-3/2$) is suppressed by a factor of $\tfrac{m}{M_5}\ll1$ relatively to $\mu$ (with mass dimension $-1/2$), thus it is generically expected to be small. Nonetheless, for a mass $m$ near the five-dimensional Planck scale, it can be important.

\medskip
In $d=4$, summing up the leading contribution to \eqref{be_quartic_4}, \eqref{be_photon_4}, \eqref{be_graviton_4},\eqref{be_scalar_massless_4}, we obtain:
\begin{align}
\sum_i q_i^2  g_i^2  - \frac{1}{2} m^2 \kappa^2 - \bigg(  \mu -  \frac{m\delta}{2} \bigg)^2 
+ \frac{1}{2L} \left( \frac{a}{m} - m(b-2c) \right)
\ge 0
\,,
\label{flat_limit_4}
\end{align}
where we used that all $p=4,q=3$ and $p=3,q=2$ hypergeometric functions in \eqref{hypergeometrics} satisfy $\lim_{\Delta\to\infty}\,_4 F_3=4 (-1 + \sqrt{2}) \Delta$ and $\lim_{\Delta\to\infty}\,_3 F_2=\calo(\Delta^{-1/2})$ respectively. Thus, only $\,_4 F_3$'s contribute to \eqref{flat_limit_4}, and under the same suppressions discussed above, we reproduce the four-dimensional flat space expression with the right coefficients.

	\subsection{BPS states}
	\label{sec:BPS}
	
In this Section we illustrate two explicit examples where a BPS state $\phi$ saturates the self-binding energy $\gamma=0$. The first example is taken from the $\caln=2$, $d=5$ gauged supergravity described in Section 4.2 of \cite{Ceresole:2001wi}. The second example is taken from the $\caln=2$, $d=4$ gauged supergravity described in Section 4.3 of \cite{Hristov:2009uj}. 
We will quickly review the setups in our notation, and refer the reader to the sources for a more detailed treatment. 

\subsubsection{Supersymmetric $\ads_5$}
\label{sec:BPS5}

The (bosonic sector of the) gauged supergravity we are interested in consists of a gravity multiplet, containing graviton and graviphoton $A_0$, a hypermultiplet, containing four real scalars $q^X=(V,\sigma,\theta,\tau)$, and a vector multiplet, containing a real scalar $\rho$ and an additional vector $A_1$. 
The moduli space is the product of a quaternionic manifold $\calm_Q=\tfrac{SU(2,1)}{SU(2)\times U(1)}$ (parametrized by $q^X$) and a special K\"ahler manifold $\calm_{SK}=O(1,1)$ (parametrized by $\rho$), and we will gauge $q^X$ under $U(1)\times U(1)$ as explained below. The action is (setting $\kappa=1$):
\begin{align}
\frac{\call}{\sqrt{-g}} 
&= 
\frac{R}{2} 
- \frac{1}{4} a_{00} F_0^2 
- \frac{1}{4} a_{11} F_1^2
- \frac{1}{2} g_{XY} D_\mu q^X D^\mu q^Y
- \frac{1}{2} g_{\rho\rho} \del_\mu \rho \del^\mu \rho
- V(q,\rho)
\,,
\end{align}
up to a Chern--Simons term that is not relevant to our analysis. The covariant derivative is $D_\mu q^X = \del_\mu q^X + g A_\mu^0 K_0^X(q) + g A_\mu^1 K_1^X(q)$, where 
$K_0^X(q),K_1^X(q)$ are the Killing vectors of the gauged isometries on $\calm_Q$, and
\begin{align}
a_{00} = \frac{1}{\rho^8}
\,,
\quad
a_{11} = \rho^4
\,,
\quad
g_{\rho\rho} = \frac{12}{\rho^2}
\,,
\end{align}
while the hyperscalar metric $g_{XY}$ is
\begin{align}
\d s^2 = 
\frac{\d V^2}{2V^2} 
+ \frac{1}{2V^2} (\d\sigma+2\theta\d\tau - 2\tau\d\theta)^2
+ \frac{2}{V} (\d\tau^2 + \d\theta^2)
\,.
\end{align}
The hyperscalars are charged under the abelian subgroup 
$U(1)\times U(1) \subset SU(2)\times U(1) $ via the following Killing vectors
\begin{align}
\label{Killing}
\nn
K_0 &= 
\sqrt{2} \alpha \left( \frac{T_3}{2} + \frac{\gamma}{\sqrt{3}} T_8  \right)
\,, \\
K_1 &= 
\alpha \left( T_3 + \frac{\beta}{\sqrt{3}} T_8  \right)
\,,
\end{align}
where $\alpha,\beta, \gamma$ are real ($\alpha>0$ without loss of generality) and $T_3,T_8$ are the $U(1)^2$ generators, explicitly:
\begin{align}
\label{generators}
T_3 = 
\begin{pmatrix}
\frac{\sigma V}{2} \\
\frac{1}{4} \left(\sigma ^2-\left(\theta^2+\tau^2+V\right)^2+1\right) \\
\frac{1}{4} \left(\theta\sigma-\tau^3-\tau \left(\theta^2+V-3\right)\right) \\
\frac{1}{4} \left(\theta ^3+\sigma \tau +\theta  \left(\tau^2+V-3\right)\right)
\end{pmatrix}
\,, 
\quad
T_8 = 
\begin{pmatrix}
\frac{1}{2} \sqrt{3} \sigma  V    \\
\frac{1}{4} \sqrt{3} \left(\sigma^2-\left(\theta ^2+\tau^2+V\right)^2+1\right) \\ 
-\frac{1}{4} \sqrt{3} \left(-\theta  \sigma +\tau^3+\tau  \left(\theta^2+V+1\right)\right)  \\
\frac{1}{4} \sqrt{3} \left(\theta^3+\sigma  \tau +\theta \left(\tau^2+V+1\right)\right) 
\end{pmatrix}
\,.
\end{align}
The superpotential is 
\begin{align}
\label{superpotential_W}
W =  \sqrt{\frac{2}{3} \vec P \cdot \vec P}
\,, \quad
\vec P = \frac{\alpha}{3\sqrt{2} \rho^2}
\big(\sqrt{3}(\rho^6+2)\vec P_3+2\vec P_8(\beta+\gamma\rho^6)\big)
\,,
\end{align}
with
\begin{align}
\nn
\vec P 
=
\frac{\sqrt{2} \vec P_1 + \vec P_0 \rho^6}{\sqrt{3}\rho^2}
= \frac{\alpha}{3\sqrt{2} \rho^2}
\big(\sqrt{3}(\rho^6+2)\vec P_{T_3}+2\vec P_{T_8}(\beta+\gamma\rho^6)\big)
\,,
\end{align}
where we used that $\vec P_0,\vec P_1$ are the prepotentials corresponding to $K_0,K_1$, namely
$\vec P_0 = \sqrt{2} \alpha \left( \frac{1}{2} \vec P_{T_3} + \frac{\gamma}{\sqrt{3}} \vec P_{T_8}  \right)$ and 
$\vec P_1 = \alpha \left(\vec P_{T_3} + \frac{\beta}{\sqrt{3}} \vec P_{T_8}  \right)$. 
Prepotentials (or moment maps) $\vec P_{T_3},\vec P_{T_8}$ are
\begin{align}
\nn
& \vec P_{T_3} = 
\left(
\begin{matrix}
-\frac{\theta ^3+\sigma \tau +\theta  \left(\tau^2-V-3\right)}{4\sqrt{V}} \\ 
\frac{-\theta ^2\tau+\theta\sigma+\tau  \left(-\tau^2+V+3\right)}{4\sqrt{V}} \\
-\frac{\left(\theta^2+\tau^2\right)^2-6\theta^2+\sigma^2-6 \tau^2+V^2-6 V \left(\theta^2+\tau^2-1\right)+1}{16V}
\end{matrix}
\right)
\,, \\
& \vec P_{T_8} = 
\left(
\begin{matrix}
-\frac{\sqrt{3}\left(\theta^3+\sigma\tau+\theta  \left(\tau^2-V+1\right)\right)}{4\sqrt{V}}  \\
-\frac{\sqrt{3} \left(\theta^2 \tau -\theta\sigma +\tau^3+\tau -\tau V\right)}{4\sqrt{V}} \\ 
-\frac{\sqrt{3} \left(\theta ^4+2 \theta^2 \left(\tau^2+1\right)+\sigma^2+\left(\tau^2+1\right)^2+V^2-2 V\left(3 \theta ^2+3 \tau^2+1\right)\right)}{16V}
\end{matrix}
\right)
\,.
\label{prepotentials}
\end{align}
The scalar potential is obtained from the superpotential as
\begin{align}
\label{potential_V}
V = g^2 
\left( 
- 6 W^2 
+ \frac{9}{2} g^{\rho\rho} \del_\rho W \del_\rho W
+ \frac{9}{2} g^{XY} \del_X W \del_Y W
\right)
\,.
\end{align}
Supersymmetric vacua are found solving $\del_\rho W = \del_X W = 0$.\footnote{In order to simplify computations, one can work in $w=W^2$, then second and third terms in \eqref{potential_V} become $ g^{\rho\rho} \frac{\del_\rho w \del_\rho w}{4w} +  g^{XY} \frac{\del_X w \del_Y w}{4w}$ and AdS susy vacua are found satisfying $\del_\rho w = \del_X w = 0$ with $w \neq 0$.}
This yields two types of SUSY AdS vacua, an isolated point and a circle of points (called critical point 1 and 2 in \cite{Ceresole:2001wi} respectively). Here, we will study the effective action for fluctuations around the isolated point, which has the following vevs
\begin{align}
(V,\sigma,\theta,\tau) = (1,0,0,0) 
\,, \quad
\rho = 1
\,.
\end{align}
One can see that Killing vectors \eqref{Killing} vanish in this vacuum, thus $A_0,A_1$ are massless. 
More conveniently, we use $\rho=e^\chi$ and complex coordinates $\phi_1$, $\phi_2$ defined by
\begin{align}
\label{newcoords}
&V = \frac{1-|\phi_1|^2-|\phi_2|^2}{(1+\phi_1)(1+\phi^*_1)} 
\,, \quad
\sigma = i \frac{\phi_1-\phi^*_1}{(1+\phi_1)(1+\phi^*_1)} 
\,, \\
& \theta = \frac{\phi_2}{2(1+\phi_1)} + \frac{\phi_2^*}{2(1+\phi^*_1)} 
\,, \quad
\tau = \frac{i \phi_2}{2(1+\phi_1)} - \frac{i \phi_2^*}{2(1+\phi^*_1)} 
\,,
\end{align}
with vevs
\begin{align}
\label{vacuum1}
\phi_1 = \phi_2 = \chi = 0 
\,.
\end{align}
Expanding kinetic terms and potential around this vacuum, canonicalizing fluctuations, and fixing $\alpha$ such that $\Lambda=-\frac{6}{L^2}$, we obtain an effective action for fluctuations of the form \eqref{setup_d} (reintroducing $\kappa$):
\begin{align}
\nn
& \frac{\call}{\sqrt{-g}} 
= 
- \frac{F_0^2}{4} 
- \frac{F_1^2}{4} 
- |D_\mu \phi_1|^2 - |D_\mu \phi_2|^2 
- \frac{(\del\chi)^2 }{2}
+ \frac{6}{\kappa^2 L^2} 
- m_1^2 |\phi_1|^2 
- m_2^2 |\phi_2|^2 
- \frac{M^2}{2} \chi^2
\\
& ~~~ 
- Y_1 \chi |\phi_1|^2 
- Y_2 \chi |\phi_2|^2 
- a_1 |\phi_1|^4
- a_2 |\phi_2|^4 
- b_1 |\phi_1|^2 |\del_\mu \phi_1|^2
- b_2 |\phi_1|^2 |\del_\mu \phi_1|^2
+ \dots
\,,
\label{sugra_eff}
\end{align}
where $D_\mu \phi_i = \del_\mu \phi_i - i \big( g_0 q_{i 0} A_{0\mu} + g_1 q_{i 1} A_{1\mu} \big) \phi_i$, $\delta=c=0$ and
\begin{align}
\nn
& g_0 q_{10} = 
\frac{\kappa}{\sqrt{2} L} (2\gamma+1) 
\,, 
\quad
g_1 q_{11} = 
\frac{\kappa}{L} (\beta+1) 
\,,\quad
g_0 q_{20} 
=
\frac{\kappa}{\sqrt{2} L} (2\gamma-1)
\,, 
\quad
g_1 q_{21} 
=
\frac{\kappa}{L} (\beta-1)
\,, 
\\
\nn
& Y_1 = 
- \frac{\kappa}{\sqrt{3}L^2} (\beta-2\gamma)  (1+2\beta+2\gamma) 
\,, 
\quad
Y_2 = 
- \frac{\kappa}{\sqrt{3}L^2} (\beta-2\gamma) (-1+2\beta+2\gamma) 
\,, \\
\nn
&  m_1^2 = 
\frac{1}{4L^2 } \left(-5+2(\beta+\gamma)\right) \left(3+2(\beta+\gamma)\right) 
\,,
\quad 
m_2^2 =  
\frac{1}{4L^2 } \left(5+2(\beta+\gamma)\right) \left(-3+2(\beta+\gamma)\right)
\,, \\ 
&  M^2 = 
- \frac{4}{L^2}
\,, 
\quad 
a_1 = a_2 =  \frac{\kappa^2}{2L^2} (-6+3\beta^2+4\beta\gamma+4\gamma^2) 
\,, 
\quad 
b_1 = b_2 = 2 \kappa^2
\,,
\label{coeffs}
\end{align}
while dots contain all the other terms, like the Chern–Simons term and higher order terms ($\phi_1^2$, $\phi_2^2$, $\chi^3$, $\chi^4$, $\chi^2 \phi_i^2$, etc.), that we can neglect in our analysis. 
Notice that $m_1^2-m_2^2$ is proportional to  $-(\beta+\gamma)$. If $\beta+\gamma>0$, $\phi_1$ is associated to a primary operator and $\phi_2$ to its dual. If $\beta+\gamma<0$, the viceversa is true. In any case, the primary has conformal dimension $\Delta=\frac{3}{2}+|\beta+\gamma|>\frac{3}{2}$ and its dual has dimension $\Delta+1>\frac{5}{2}$. In the following we take $\beta+\gamma>0$, and thus $\phi_1$ is primary.

\medskip
Since the effective action \eqref{sugra_eff} matches \eqref{setup_d} with $\beta=0$, we can readily compute the self-binding energy for the primary $\phi_1$ by plugging parameter values \eqref{coeffs} into contributions \eqref{be_quartic}, \eqref{be_photon}, \eqref{be_graviton} and \eqref{be_BF}, thus obtaining:
\begin{align}
\label{be_BPS_ab}
\gamma^{V} &= 
\frac{\kappa^2}{L^4}
\frac{\pi ^2 (\beta +1) (3 \beta -4 \Delta +3) N_{\Delta }^4}{2 \left(2 \Delta ^2-3 \Delta +1\right)}
\,, \\
\label{be_BPS_A0}
\gamma^{A_0} &= 
\frac{\kappa^2}{L^4} \frac{2 \pi ^2 (\beta -\Delta +1)^2 N_{\Delta }^4}{(2\Delta -1)}
\,, \\
\label{be_BPS_A1}
\gamma^{A_1} &= 
\frac{\kappa^2}{L^4} \frac{\pi ^2 (\beta +1)^2 N_{\Delta }^4}{(2 \Delta-1) }
\,, \\
\label{be_BPS_h}
\gamma^{\rm grav} &= 
- \frac{\kappa^2}{L^4} \frac{2 \pi ^2 (\Delta -2) \Delta ^2 \kappa ^2 N_{\Delta }^4}{3 (\Delta -1) (2 \Delta -1) }
\,, \\
\label{be_BPS_chi}
\gamma^{\rm scal} &= 
- \frac{\kappa^2}{L^4}
\frac{\pi ^2 (3 \beta -2 \Delta +3)^2 N_{\Delta}^4}{6 (\Delta -1)}
\,, 
\end{align}
where we used $\gamma=\Delta-\frac{3}{2}-\beta$. A little algebra shows that the total self-binding energy $\gamma_{\phi_1\phi_1} = \gamma^{V} + \gamma^{A_0} + \gamma^{A_1} + \gamma^{\rm grav}  + \gamma^{\rm scal} = 0 $ as expected for a BPS state. On the other hand, one can check that the self-binding energy for $\phi_2$, $\gamma_{\phi_2\phi_2}$ can be positive, negative or vanishing, depending on the value of gauging parameters $\beta,\gamma$.

\medskip
Our result also reproduces the one obtained in Section 5 of \cite{Fitzpatrick:2011hh}, as the sub-case where a photon and $\chi$ decouple. This situation is reproduced for $\beta = 2\gamma > 0$ (where $\beta=c$ in \cite{Fitzpatrick:2011hh}) such that 
\begin{align}
\label{Shih_subcase}
\Delta=\frac{3}{2}(\beta+1)
\,,
\quad
Y_1=0 
\,,
\quad
g_0^2 q^2_{10} + g_1^2 q_{11}^2 = \frac{3\kappa^2}{2L^2}(\beta+1)^2
\,,
\quad
a_1=\frac{3}{L^2}(\beta^2-1)
\,.
\end{align}
We thus find again $\gamma_{\phi_1\phi_1} = 0 $. In particular, the contributions from the graviton and the potential $V$ are the same as in \cite{Fitzpatrick:2011hh} while $\gamma^{A_0} + \gamma^{A_1} = \gamma^{B_0}$ with $B_0$ the single photon with $gq=\tfrac{\sqrt{3}}{\sqrt{2}}\tfrac{\kappa}{L}(\beta+1)$ as in \cite{Fitzpatrick:2011hh}. 
With a little effort, this can be clearly understood at the level of effective lagrangian. We can indeed match \eqref{sugra_eff} to eq.~5.13 in \cite{Fitzpatrick:2011hh} by rotating our $U(1)^2$ basis to the new basis given by $B_0=-\tfrac{1}{\sqrt{3}}(A_0+\sqrt{2}A_1)$ and $B_1=-\tfrac{1}{\sqrt{3}}(\sqrt{2}A_0-A_1)$. 
The resulting lagrangian corresponds to eq.~5.13 in \cite{Fitzpatrick:2011hh}, where $\phi_1,\phi_2$ (respectively $z_2, z_1$ there \footnote{Notice that $z_2$ is primary in \cite{Fitzpatrick:2011hh}.}) are charged under $B_0$ only, with the additional presence of a free (ungauged) photon $B_1$ and a neutral scalar $\chi$ which is sourced by quartic terms (since $Y_1=0$). As a consequence, neither $B_1$ nor $\rho$ contribute to the total binding energy for $\phi_1$.

	
\subsubsection{Supersymmetric $\ads_4$}
\label{sec:BPS4}

A simple, non-trivial example we can work out in $d=4$ $\caln=2$ gauged supegravity is described in Section 4.3 of \cite{Hristov:2009uj}, and we will follow this reference. In its simplest realization, it consist of the gravity multiplet, with graviphoton $A$, and a single hypermultiplet, with four real scalars $q^X=(\rho,\sigma,\xi_1,\xi_2)$.\footnote{In the notation of \cite{Hristov:2009uj}, $\xi = \xi_1 + i \xi_2$.} The moduli space is just the quaternionic manifold $\calm_Q=\tfrac{SU(2,1)}{SU(2)\times U(1)}$ with metric ($\rho>0$)
\begin{align}
\d s^2 = 
\frac{\d \rho^2}{4 \rho^2} 
+ \frac{1}{4\rho^2} (\d\sigma + 2\xi_2\d\xi_1 -  2\xi_1\d\xi_2 )^2
+ \frac{1}{\rho} (\d\xi_1^2 + \d\xi_2^2)
\,.
\end{align}
and we gauge (part of) $q^X$ under the graviphoton. 
The action is (setting $\kappa=1$) \cite{Guarino:2017jly}:
\begin{align}
\label{sugra_4d}
\frac{\call}{\sqrt{-g}} 
&= 
\frac{R}{2} 
- g_{XY} D_\mu q^X D^\mu q^Y
- \frac{1}{8} F^2
- V(q)
\,,
\end{align}
with covariant derivative $D_\mu q^X = \del_\mu q^X - g A_\mu K^X(q)$. The Killing vector is
\begin{align}
K = \alpha (0,0,-\xi_2,\xi_1)
\,,
\label{Killing_BPS4}
\end{align}
and the corresponding moment map is
\begin{align}
\vec P = \alpha \left( \frac{2\xi_1}{\rho^{1/2}},-\frac{2\xi_2}{\rho^{1/2}},1-\frac{\xi_1^2 + \xi_2^2}{\rho} \right)
\,.
\end{align}
In absence of vector multiplets, the potential contains only two terms:
\begin{align}
V = g^2 
\left( 
4 K^X K^Y g_{XY} 
- 3 \vec P \cdot \vec P
\right)
\,.
\end{align}
$\ads$ supersymmetric vacua are obtained by requiring $K^X=0$ and $\epsilon^{ijk} P^j P^k=0$ with $P \cdot \vec P>0$.  These conditions are met for vanishing scalar vevs, $\xi_1=\xi_2=0$, and any $\rho=\rho_0>0, \sigma=\sigma_0$. Since in this vacuum $K=0$, then $A$ is massless. 
Henceforth, we will use $\rho=e^\chi$ for convenience.
Expanding around this vacuum, canonicalizing fluctuations and fixing $\alpha=\tfrac{1}{gL}$ such that $\Lambda=-\frac{3}{L^2}$, we obtain an effective action (reintroducing $\kappa$):
\begin{align}
\nn
& \frac{\call}{\sqrt{-g}} 
= 
- \frac{F^2}{4} 
- |D \phi|^2 
- \frac{(\del\chi)^2 }{2}
- \frac{(\del\sigma)^2 }{2}
+ \frac{3}{\kappa^2 L^2} 
+ \frac{2}{L^2} |\phi|^2 
\\
& ~~~ 
+ \sqrt{2} \kappa \chi |\del \phi|^2
- \frac{2\sqrt{2}\kappa}{L^2} \chi |\phi|^2 
- \frac{\kappa^2}{L^2} |\phi|^4
+ \frac{\kappa^2}{4} (\phi^\dag \del \phi - \phi \del \phi^\dag)^2
+ \dots
\,,
\label{sugra_eff_4}
\end{align}
where $D_\mu \phi = \del_\mu \phi - i \sqrt{2} \tfrac{\kappa}{L} A_{\mu} \phi$ and we have neglected all terms that do not produce exchange vertices like those of Fig.~\ref{fig:graphs}.\footnote{In particular, there is also a term $\sim i \del^\mu\sigma(\phi\del_\mu\phi^\dag-\phi^\dag\del_\mu\phi)$ in \eqref{sugra_eff_4}. As explained in footnote \ref{foot:extraterms}, integrating by parts and using the equation of motion for $\phi$, this term corresponds to a total derivative.} We immediately see that the effective action \eqref{sugra_eff_4} corresponds to \eqref{setup_d} where the non-vanishing coefficients are
\begin{align}
m^2 L^2 = -2 
\,, \quad
g^2q^2 = 2 \frac{\kappa^2}{L^2} 
\,, \quad
a = \frac{\kappa^2}{L^2}
\,, \quad
b = -2c = \frac{\kappa^2}{2}
\,, \quad
Y = \frac{2\sqrt{2}\kappa}{L^2}
\,, \quad
\beta = - \sqrt{2} \kappa
\,.
\label{parameters_BPS4}
\end{align}
From $m^2 L^2 = \Delta(\Delta-3)$ and the requirement $\Delta\ge\tfrac{3}{2}$ (see after \eqref{massless_sol_4}), we deduce $\Delta=2$, and thus the binding energies \eqref{be_quartic_4},\eqref{be_photon_4},\eqref{be_graviton_4},\eqref{be_scalar_massless_4} are
\begin{align}
\label{bes_5}
\gamma^{V} = 
\gamma^{\rm scal} =
- \frac{3}{8} \frac{\kappa^2}{L^3} \pi ^2 N_{2}^4
\,, \qquad
\gamma^{\rm phot} = 
\frac{5}{4} \frac{\kappa^2}{L^3} \pi ^2 N_{2}^4
\,, \qquad
\gamma^{\rm grav} = 
- \frac{1}{2} \frac{\kappa^2}{L^3} \pi ^2 N_{2}^4
\,,
\end{align}
and the total binding energy is $\gamma_{\phi\phi}=0$, as expected. 
As described in Section 4.3 of \cite{Hristov:2009uj}, this example can be generalized to include a vector multiplet containing a complex scalar and an additional vector potential $A_1$. 
As one can check, the relevant part of the resulting effective action around the supersymmetric AdS vacuum is almost identical, since the (two real) scalars in the vector multiplet do not furnish couplings we are interested in. The only difference is that the $U(1)$ force is now split between two photons $A_0, A_1$ in a way that $g_0^2 q_0^2+g_1^2 q_1^2=2 \tfrac{\kappa^2}{L^2} $, meaning that the total gauge force is the same as in \eqref{parameters_BPS4}. Thus the same conclusion holds.

	\section{Summary}
	\label{sec:summary}
	
	In this paper we studied the self-binding energy for a charged particle in AdS. We calculated the contribution to the binding energy from all the relevant contact terms, from the graviton, from multiple photons, and from a neutral scalar field.
	
	We are motivated by the idea that the formulation of the Weak Gravity Conjecture in AdS is the Positive Binding Conjecture of \cite{Aharony:2021mpc}. Our results translate this constraint into constraints on the effective theory coupled to gravity in AdS. 
	
	One application of our results is to testing positive binding in String theory. String theory tests are somewhat complicated by the absence of scale separation between the internal dimensions and the AdS scale in most known cases. Nonetheless, if the extra dimensions have an associated isometry the Kaluza-Klein modes carry charge under it and therefore cannot contribute to self-binding at tree-level. The binding energy would then be determined by the zero modes only. Interestingly, it may be that scale separation itself may be related to the WGC, for example as in \cite{Cribiori:2022trc}, and it would be interesting to study if there can be a connection with positive binding.  
	
	Our results applied to charged scalar fields playing the role of the particle with positive binding. It would be interesting to generalise them to fermions, or higher-spin fields. Another interesting direction is to consider states with a large angular momentum. These were studied through the CFT dual in \cite{Cuomo:2022kio} and shown to have (asymptotically) negative binding energy. It would be good to understand the transition from positive binding to negative binding at large angular momentum from the gravity side. 
	
	A general point of this paper is that the precise formulation of Swampland constraints, in this case the WGC or Repulsive Force Conjecture, can qualitatively differ in AdS from flat space. It would be interesting to repeat this type of analysis for other similar Swampland constraints that were formulated in flat space. For example, versions of the WGC applied not to point particles (such as axions and higher dimension objects, see \cite{Palti:2019pca} for a review), or the proposed constraint on bound state sizes \cite{Freivogel:2019mtr}. 
	
	\vskip 20pt
	\noindent {\bf Acknowledgements:} We thank Ofer Aharony for extremely useful discussions and explanations. We also thank Nicol\`o Petri, Niccol\`o Cribiori, Arthur Hebecker, Gerben Venken, Timo Weigand and Enrico Andriolo for other useful discussions. 
	The work in this paper was supported by the Israel Science Foundation (grant No. 741/20) and by the German Research Foundation through a German-Israeli Project Cooperation (DIP) grant ``Holography and the Swampland". The work of MM was supported by a Minerva Fellowship of the Minerva Stiftung Gesellschaft f\"ur die Forschung mbH.

\appendix

\section{Quantization of a free scalar in AdS}
\label{app:free}

We will work in Lorentzian signature and follow the detailed procedure described in \cite{Balasubramanian:1998sn}, taking $\ads_5$ \eqref{AdS_global} (with $L=1$) for concreteness. The action for a free complex scalar $\phi$ of mass $m$,
\begin{align}
\label{scalar_action}
S[\phi] = - \int_{\ads_5} \d^5x \sqrt{-g} \left( |\del\phi|^2 + m^2|\phi|^2 \right)
\,,
\end{align}
yields the following equation of motion
\begin{align}
\label{free_eom}
\frac{1}{\sqrt{-g}} \del_\mu ( \sqrt{-g} \del^\mu \phi) - m^2 \phi = 0 
\,.
\end{align}
Here $\phi$ is fixed at initial and final equal time hypersurfaces $\Sigma_i,\Sigma_f$ (at times $t_i,t_f$ respectively), as well as at the AdS boundary at $y=0$, $\delta\phi_{y=0}=0$.

In order to canonically quantizatize $\phi$ in the Hamiltonian formalism, we need to find a complete, normalizable set of modes, which in AdS this corresponds to choosing some sort of boundary condition at $y=0$.
As suggested by Breitenlohner and Freedman (BF), a meaningful boundary condition is to require that energy and the Klein--Gordon product \eqref{KGprod} are conserved, \cite{Breitenlohner:1982bm, Breitenlohner:1982jf}. The Klein--Gordon (KG) product is the inner product adopted to define norms. Given a space-like slice $\Sigma$, it is defined as 
\begin{align}
\label{KGprod}
(\phi_1,\phi_2) \equiv 
i \int_\Sigma \d^4 x \sqrt{-g} g^{tt} \phi_1^* (\overleftarrow{\del_t} - \overrightarrow{\del_t}) \phi_2
\,.
\end{align}
As noticed in \cite{Balasubramanian:1998sn} and reviewed below, it is sufficient to require mode normalizability in order to identify the correct modes; the BF consideration turns out to be relevant in a special case where normalizability alone is not conclusive.  

\medskip
Modes are found using a Fourier decomposition on four-dimensional hypersurfaces at fixed $y$: $$\psi_{\omega lJ} = e^{-i \omega t} \caly_{lJ}(\Omega) f_{\omega lJ}(y) \,,$$
where $\caly_{lJ}(\Omega)$ are normalized eigenstates of the Laplacian on $S^3$ with eigenvalue $-l(l+2)$ and $J$ labels all other angular quantum numbers. 
Plugging this Ansatz into equation \eqref{free_eom}, we obtain an equation for modes:
\begin{align}
\frac{y^3}{1-y^2} \del_y \left( \frac{(1-y^2)^2}{y^3} \del_y f_{\omega lJ} \right)
+ \left( \omega^2 - \frac{l(l+2)}{1-y^2} - \frac{m^2}{y^2} \right) 
f_{\omega lJ}(y) = 0
\,.
\end{align}
By requiring $f_{\omega lJ}$ to be smooth at the origin $y=1$, we find that the solution can be expressed in terms of Gauss hypergeometric functions ${}_2F_1$ (forgetting the normalization for the moment) as
\begin{align}
\label{modes1}
f_{\omega lJ} = (1-y^2)^{l/2} y^{\Delta_+} \, {}_2F_1 \left(\tfrac{1}{2} (\Delta_++l-\omega),\tfrac{1}{2} (\Delta_++l+\omega),l+2,1-y^2 \right)
\,, 
\end{align}
where $\Delta_+$ is the largest root of $\Delta(\Delta-4)=m^2$,
\begin{align}
\Delta_{\pm} = 2 \pm \nu \,, 
\quad
\nu \equiv \sqrt{4+m^2}
\,.
\end{align}
The discussion regarding boundary behaviour, quantization and normalization of the solution depend on whether $\nu$ is integer or not.

\subsection*{The case with $\nu\not\in\mathbb{Z}$}

We can use hypergeometric identities to recast \eqref{modes1} as\footnote{\eqref{modes1} is $\Psi^{(1)}$ in the language of \cite{Balasubramanian:1998sn} (see Sect.~3.2 there), and \eqref{modes2} corresponds to eq.~(36) there.}
\begin{align}
\label{modes2}
f_{\omega lJ} = C_+ \Phi^+_{\omega lJ} + C_-  \Phi^-_{\omega lJ} 
\,,
\end{align}
where 
\begin{gather}
\nonumber
\Phi^\pm_{\omega lJ} = 
(1-y^2)^{l/2} y^{\Delta_\pm} \, {}_2F_1 \left(\tfrac{1}{2} (\Delta_\pm +l-\omega),\tfrac{1}{2} (\Delta_\pm +l+\omega),\Delta_\pm-1,y^2 \right)
\,, \\
\nonumber
C_+ = \frac{\Gamma(l+2)\Gamma(-\nu)}{\Gamma\left(\tfrac{1}{2} (\Delta_- +l-\omega)\right)\Gamma\left(\tfrac{1}{2} (\Delta_- +l+\omega)\right)}
\,, \\
C_- = 
\frac{\Gamma(l+2)\Gamma(\nu)}{\Gamma\left(\tfrac{1}{2} (\Delta_+ +l-\omega)\right)\Gamma\left(\tfrac{1}{2} (\Delta_+ +l+\omega)\right)}
\,,
\label{Phi_+-}
\end{gather}
Notice that $\Phi^\pm_{\omega lJ}$ have the following boundary behaviour
\begin{align}
\Phi^\pm_{\omega lJ} \sim y^{\Delta_\pm} + \calo(y^2)
\,,
\quad y\to 0
\,.
\end{align}

For $\nu>1$ (i.e., $\Delta_-<1$ and $\Delta_+>3$), $C_-$ must vanish since $\Phi^-_{\omega lJ}$ is non-normalizable (its norm diverges at the boundary).\footnote{Using \eqref{KGprod}, it is indeed easy to check that normalizability requires a fall-off faster than $y$ at the boundary.} This can happen only if one of the gamma functions at the denominator has zero or a negative integer as its argument. Remembering that we are interested in solution with positive energy, $\omega>0$, we obtain the quantization condition
\begin{align}
\label{quantization+}
\omega = \Delta_+ + l + 2n  
\,, \quad
n = 0,1,2,\dots \,.
\end{align}
If we require $\omega$ to be real, we need $\Delta$ to be real. That is, the mass has to be bounded as
\begin{align}
\label{BF_bound}
m^2\ge-4
\,,
\end{align}
This is also known as the BF bound, originally discussed by Breitenlohner and Freedman \cite{Breitenlohner:1982bm,Breitenlohner:1982jf}. 
When the BF bound is saturated, $\Delta_+=\Delta_-=2$ ($\nu=0$), and we will discuss this particular case below. 
One can check that these modes satisfy the BF requirements on conservation of energy and KG product.

\medskip
For $\nu<1$ (i.e., $1<\Delta_-<2$ and $2<\Delta_+<3$) both $\Phi^\pm_{\omega lJ}$ are normalizable. 
In this situation, Breitenlohner and Freedman have showed that conservation of energy and Klein--Gordon product force us to choose \emph{either} modes, but not both. We can choose to quantize as \eqref{quantization+} and keep only $\Phi^+_{\omega lJ}$ or alternatively quantize as
\begin{align}
\label{quantization-}
\omega = \Delta_- + l + 2n  
\,, \quad
n = 0,1,2,\dots \,.
\end{align}
and keep $\Phi^-_{\omega lJ}$. 

\subsection*{The case with $\nu=0,1,2,\dots$}

In this case, we can recast modes \eqref{modes1} as a linear combination of $\Phi^+_{\omega lJ}$ and another function, $\tilde \Phi^-_{\omega lJ}$ ($\neq\Phi^-_{\omega lJ}$ in \eqref{Phi_+-}):
\begin{itemize}
\item For \textbf{$\nu>0$}, $\tilde \Phi^-_{\omega lJ}$ blows up at the boundary, and thus one has to kill these modes by quantizing as \eqref{quantization+}. One is left with normalizable modes falling off as $y^{\Delta_+}$.
\item For $\nu=0$ (saturation of the BF bound \eqref{BF_bound}), $\tilde \Phi^-_{\omega lJ}$ contains logarithmic terms but is normalizable.
However, there is only one possible way to quantize this case, since $\Delta_+=\Delta_-=2$, and this corresponds to killing log terms. 
This can be explicitly seen by setting $\Delta=2,\nu=0$ in \eqref{modes1} and then expanding it towards the boundary:
\begin{align}
f_{\omega lJ} \sim 
\Gamma(l+2) 
\frac{H_{-n-1} + H_{n+l+1} + 2\log(y)}{\Gamma(-n)\Gamma(n+l+2)} 
y^2 
\,, \quad
y\to 0
\,,
\end{align}
where $H_n \equiv \sum_{k=1}^n \frac{1}{k}$ are harmonic numbers. Sub-leading terms have similar behaviour. 
Thus, by imposing the quantization $n=0,1,2,\dots$, one finds that the fraction simplifies to $-\frac{n!}{\Gamma(n+l+2)}$ and thus the log dependence drops out. The same happens in the sub-leading terms. The boundary behaviour is therefore a fall-off $\sim y^2$.
\end{itemize}
Summarizing, the generic mode expansion is 
\begin{align}
\label{mode_exp}
\phi = \sum_{nlJ} \left(a_{nlJ} \psi_{nlJ} + b^*_{nlJ} \psi^*_{nlJ} \right)
\,,
\end{align}
with orthonormal modes 
\begin{align}
\psi_{nlJ} &= 
N_{\Delta nl} \, 
e^{-i \omega t}\, 
\caly_{lJ}(\Omega) 
(1-y^2)^{l/2} y^{\Delta_+} \, {}_2F_1 \left(\tfrac{1}{2} (\Delta_++l-\omega),\tfrac{1}{2} (\Delta_++l+\omega),l+2,1-y^2 \right) \,, 
\label{modes}
\end{align}
where the quantization condition $\omega = \Delta_\pm + 2n + l$, $n=0,1,\dots$ depends on $\nu$ as discussed above. $N_{\Delta nl}$ is the overall normalization constant determined by the orthonormality relation,
\begin{align}
\label{orthonormality}
(\psi_{nlJ} , \psi_{n'l'J'})
=\delta_{n n'} \delta_{l l'} \delta_{J J'}
\,,
\end{align}
and coefficients in \eqref{mode_exp} are fixed by initial data $\phi(t_i,y,\theta_1,\theta_2,\varphi)$ and $\partial_t \phi(t_i,y,\theta_1,\theta_2,\varphi)$ using the KG product. 

Canonical quantization proceeds \eqref{mode_exp} as usual by promoting coefficients to creation and annihilation operators 
\begin{align}
\phi = \sum_{nlJ} \left(a_{nlJ} \psi_{nlJ} + b^\dag_{nlJ} \psi^*_{nlJ} \right)
\,,
\end{align}
with canonical commutation relations
\begin{align}
[a_{nlJ} , a_{n'l'J'}^\dagger]
=[b_{nlJ} , b_{n'l'J'}^\dagger]
=\delta_{n n'} \delta_{l l'} \delta_{J J'}
\,.
\end{align}
Given the momentum conjugate 
$\Pi_{\phi}= - g^{tt}\del_t\phi^\dag$, 
$\Pi_{\phi^\dag}= - g^{tt}\del_t\phi$, 
the free Hamiltonian density is 
\begin{align}
\label{Hdensity_free}
\calh = - g_{tt} \Pi_{\phi}\Pi_{\phi^\dag} + |\vec\nabla\phi|^2 + m^2 |\phi|^2
\,.
\end{align}
Using the orthonormality relation \eqref{orthonormality}, the free Hamiltonian $H = \int \d^{4}x \sqrt{-g} \calh$ is
\begin{align}
\label{free_Hamiltonian}
H = 
\sum_{nlJ} \omega ~ (a^\dag_{nlJ}a_{nlJ} + b^\dag_{nlJ}b_{nlJ}  )
\,, \quad
\omega = \Delta + 2n + l 
\,, \quad
n,l = 0,1,2,\dots 
\,,
\end{align}
where $\Delta$ can be either $\Delta_\pm$.
The single and two-particle states of interest here are the ones with lowest energy $n=l=J=0$,
\begin{align}
\label{ketphiphi}
\ket*{\phi} = b_0^\dag \ket*{0} \,,
\quad
\ket*{\phi\phi} = \frac{1}{\sqrt{2}} b_0^\dag b_0^\dag \ket*{0} \,,
\quad 
b_0\equiv b_{000} \,, 
\end{align}
corresponding to the eigenfunction 
\begin{align}
\label{app:psi0}
\psi_0(x) \equiv \psi_{000}(x) = N_\Delta e^{-it\Delta} y^\Delta \,,
\quad
N_\Delta = \sqrt{\frac{\Delta-1}{2\pi^2}} \,.
\end{align}
As expected in a free theory, $H \ket{\phi} = \Delta \ket{\phi}$ and $H \ket{\phi\phi} = 2\Delta\ket{\phi\phi}$. 

\medskip
Mutatis mutandis, the same quantization procedure applies in $\ads_4$, obtaining
\begin{align}
\label{app:psi0_4}
\psi_0(x) = N_\Delta e^{-it\Delta} y^\Delta \,,
\quad
N_\Delta =  \frac{1}{\pi^{3/4}} \sqrt{\frac{\Gamma(\Delta+1)}{2\Delta\Gamma(\Delta-1/2)}}
\,,
\end{align}
as the lowest energy state of interest. In $\ads_4$, $\Delta$ can be either root of $\Delta(\Delta-3)=m^2$, and thus the BF bound is $m^2\ge-\frac{9}{4}$.

	\bibliographystyle{ssg}
	\bibliography{susyswamp.bib}  
\end{document}